\documentclass[10pt,a4paper]{article}
\usepackage[utf8]{inputenc}
\usepackage[english]{babel}

\usepackage{amsfonts}
\usepackage{amssymb}
\usepackage{graphicx}

\usepackage{xcolor}
\definecolor{light-gray}{gray}{0.90}
\usepackage[plain,procnumbered,noline]{algorithm2e}
\usepackage[numbered,autolinebreaks,useliterate,framed]{mcode}
\usepackage{url}
\usepackage{caption}

\usepackage[numbers]{natbib}

\usepackage{amsmath}

\DeclareMathOperator*{\argmin}{arg\!min}
\DeclareMathOperator{\signum}{sgn}

\let\proglang=\textsf

\allowdisplaybreaks

\SetKwInOut{Input}{Input}
\SetKwInOut{Output}{Output}

\providecommand{\keywords}[1]
{
  \small	
  \textbf{\textit{Keywords---}} #1
}

\author{Ivano Azzini,  Domenico Perrotta and Francesca Torti\\[3mm] European Commission, Joint Research Centre (JRC)}
\title{A practically efficient fixed-pivot selection algorithm and its extensible MATLAB suite}
\begin{document}

\maketitle

\begin{abstract}
Many statistical problems and applications require repeated computation of order statistics, such as the median, but most statistical and programming environments do not offer in their main distribution linear selection algorithms.  
We introduce one, formally equivalent to \textsf{quickselect}, which keeps the position of the pivot fixed. This makes the implementation simpler and much practical compared with the best known solutions.
It also enables an ``oracular'' pivot position option that can reduce a lot the convergence time of certain statistical applications.
We have extended the algorithm to weighted percentiles such as the weighted median, applicable to data associated with varying precision measurements, image filtering, descriptive statistics like the medcouple and for combining multiple predictors in boosting algorithms.   
We provide the new functions in \textsf{MATLAB}, \textsf{C} and R. We have packaged them in a broad  \textsf{MATLAB} toolbox addressing robust statistical methods, many of which can be now optimised by means of efficient (weighted) selections. 
\end{abstract}


\section{Introduction}
\label{intro:sec}

The well-known \textit{selection problem}  consists in finding the \textit{$k$-th order statistic} of an unsorted array of $n \ge k$ elements, assumed distinct\footnote{
This common assumption simplifies formalism and derivations. Practical solutions also work with repeated values.} and with all their permutations equally likely. 
It is thus defined on a totally ordered set $U$ as:\\
\begin{algorithm}[H]
\vspace{1mm}
\SetKwInOut{Input}{Input}
\SetKwInOut{Output}{Output}
\Input{An array $A=(a_1 \ldots a_n)$ and an integer $k$ such that $1 \le k \le n$ and $a_i \in U$ for each $i = 1, \ldots , n$.}
\vspace{1mm}
\Output{$a_{(k)}$, the $k$-th smallest element of $A$.}
\vspace{2mm}
\end{algorithm}
\noindent
The index notation $(k)$ indicates the \textit{rank} of the $k$ element of the array $A$, which is the index in the order statistics list \( R_k(A) = \sum_{i=1}^{n} I_{[0 , \infty)}(a_k- a_i) \), being $I$ the indicator function.
The rank $(k)$ can be associated to the  statistical percentile $p_k$ by $(k) =  \lceil \frac{p_k}{100} \cdot n \rceil $, where the  ceiling function $\lceil \cdot \rceil$ gives the least integer greater than or equal to its argument. 
The \textit{median} of $A$, say $\tilde{A}$, is a particular instance of the selection problem when $n$ is odd, that is: if $n=2m +1$ for an integer $m$ then $\tilde{A}=a_{(m+1)}=a_{(\frac{n+1}{2})}$. 
Otherwise, when $n=2m$,  $\tilde{A}$ is the  arithmetic mean of the two middle order statistics $a_{(m)}=a_{(\frac{n}{2})}$ and $a_{(m+1)}=a_{(\frac{n}{2}+1)}$. 
%

The time-complexity $T(n)$ of selection algorithms is measured by counting the number of comparisons and exchanges between elements of $A$. 
The best partitioning-based methods - like the celebrated \textsf{quickselect} \citep[Hoare's \texttt{Find},][]{HoareFind:61,Hoare:71} - take on average  linear-time, 
but in the worst case $T(n)$ becomes quadratic.
Solutions that theoretically behave linearly also in the worst case \citep{BFPRT:73, DorZwick:1995} pose considerable implementation issues -- and  even take in practice  more CPU time than the naive counterparts based on sorting \citep{BleichOverton:83} -- while it is avowed that applications require more ``useful practical algorithms'' \citep[][p. 347]{SedgewickWayne:2011}. 
This paper considers an algorithm of simple implementation that satisfies this functional requirement, and we therefore named it \textsf{simpleselect}. 

The simplifications are procured thanks to an iterative fixed pivot position strategy that makes in-place array swaps around position $k$ (Section \ref{algorithm:sec}). Its efficiency is equivalent to that of \textsf{quickselect} (Section \ref{time-complexity-formal:sec})  and the chance of incurring in quadratic run-time is averted, as usual, by randomizing the initial array at a marginal cost of $\mathcal{O}(n)$ exchanges (we use backward shuffling~\cite[][pp. 124-125]{Knuth:book2full}, based on results by \cite[][pp. 26-27]{FisherYates:1948}). 
Thus, \textsf{quickselect} and \textsf{simpleselect} are equivalent ``Las Vegas'' algorithms: both produce same correct output, but while the former randomizes the pivot position at each execution step, the latter randomizes the initial array once. 

The fixed pivot and resultant simplifications enable two useful extensions. The first consists in an ``oracle'' suggesting where the order statistic value can be found in $A$. Section \ref{robstat:sec} demonstrates its benefit in two renowned robust multivariate estimators. 
The second is the extension to weighted percentiles, discussed in Section \ref{weighted:sec}.

In order to ease portability and usability, we have implemented \textsf{simpleselect} and its weighted form in \proglang{MATLAB}, \proglang{C} and \proglang{R}, which do not provide alternatives in their main distribution. 
Their filename is \texttt{quickselectFS} and \texttt{quickselectFSw}, to stress the equivalence with Hoare's algorithm.
We illustrate how to incorporate calls to the \proglang{C} function with an example in \proglang{Python} (Annex \ref{python:sec}).
To facilitate the assessment of the functions in the different environments and under general simulation settings, we have also introduced a new  \proglang{MATLAB} function that allows reproducing random numbers generated by \proglang{R} software with Mersenne Twister (Annex \ref{mtR:sec}). 

Obviously, the practical benefit of the new functions can be appreciated only in combination of other general methods relying on repeated computation of order statistics and weighted percentiles. Therefore, we have packaged them in FSDA \citep{RPT:12,RiPeCe:15}, an extensive \proglang{MATLAB} library especially addressed to robust statistics, open to contributions through \texttt{GitHub}. 
For the same reason, we have enhanced the package with new efficient functions to compute the theoretical distribution of the number of comparisons in \textsf{quickselect}-like procedures \citep{bar+pra:19} (Section \ref{vervaat:sec}) and the medcouple \citep{BrisHubertStruyf:2004} (Section \ref{medcouple:sec}), a robust skewness estimator that calls intensively the weighted median. 

We ran simulations on an Intel CPU 2.9 GHz Quad-Core i7, equipped with 16 GB RAM. We developed under \proglang{MATLAB} release R2021b, but results are consistent in much older releases.  

\section{Simpleselect}
\label{algorithm:sec}

\subsection{Concepts}
\label{concept:sec}

If the array $A$ is  ordered fully, or partially till position $k$, clearly $a_{(k)} = a_k$. But (partially) sorting the array is more than what we need if only the element $a_{(k)}$ is of interest.
%
Knuth \citep[][pp. 207–219]{Knuth:book2full} has magisterially assembled the historical roots and efforts spent to find solutions in $\mathcal{O}(n)$ time, relying on adaptations of \textsf{quicksort} \citep{Hoare:61,Geoff:86} in a \textit{divide and conquer} partitioning approach consisting of: 
\begin{enumerate}
	\item 
	A criterion to choose the position $s$ of an element $a_{s} \in A$ called \textit{pivot}.
	\item
	A procedure to rearrange the other elements of $A$ so that those in positions $1$ to $s-1$ will be smaller than those in positions $s+1$ to $n$.
\end{enumerate}	
Depending on the relative positions of the pivot and the desired order statistic, (1) and (2) are applied recursively or iteratively to one of the two parts of $A$:
\begin{itemize}
\item if $s > k$ the new target is the $k$-th element in the left side part;
\item if $s < k$ the new target is the the $(k-s)$-th element in the right side part;
\item if $s = k$ the element in position $s$ is the desired order statistic $a_{(k)}$.
\end{itemize}
In Hoare's \texttt{Find} the pivot is chosen at random. 
More complex partition-based algorithms, such as the \textsf{median of medians} \citep{BFPRT:73} and \textsf{introselect} \citep{Musser:97}, are conceptually identical, but differ for the criterion used to choose the pivot in a reasoned way -- sometimes abstruse -- in order to achieve linear worst-case.

In the following we show the practical advantages of the \textsf{simpleselect} strategy, which keeps the pivot in fixed position $k$ and iterates the swap of its value with numbers around it (right/left parts) until the pivot gets the correct value, rather than moving the pivot position (randomly or with another strategy) until it reaches the desired position $k$.
%
The performance of this simplified iterative strategy remains aligned to the best known solutions. 
Abandoning recursion also ensures scalability to large arrays, as it avoids  stack keeping issues common to most computing environments.  
%
%
We found a similar strategy, yet confined to the median computation, in \url{https://rosettacode.org/wiki/Rosetta_Code}. 
To the best of our knowledge, its properties have not been studied. 
%

\subsection{Permuting in-place}
\label{permutation:sec}

%


Finding  $a_{(k)}$ requires a permutation $\left( \sigma(1) , \ldots , \sigma(n)\right)$ of the elements of $A$ such that  $a_{\sigma(1)}, a_{\sigma(2)},  \ldots,a_{\sigma(k-1)}\le a_{\sigma(k)}$ or, equivalently, $a_{\sigma(k)}<$ $a_{\sigma(k+1)}, a_{\sigma(k+2)},  \ldots,a_{\sigma(n)}$. 
%
This is obtained with a sequence of swaps determined by the element $a_k$. 
Let us indicate with $A^t$ the status of the array at a given step $t\in\cal{N}$, with $a^t_k$  its element in position $k$, and with
 \begin{eqnarray*}
A_{L_k}^{t}  = \{ a^t_i \in A^t \; | \; a^t_i \le a^t_k   & \land \quad i < k  \} &\\
A_{R_k}^{t}  = \{ a^t_i \in A^t \; | \; a^t_i > a^t_k  & \land \quad  i > k  \} &
 \end{eqnarray*}
the sets of left and right elements of $a^t_k$ that  at step $t$  satisfy the desired ordering.
The order statistic $a_{(k)}$ is found  for some $t$ when  $\big| A_{L_k}^{t} \big| = k-1$ or, equivalently,    $\big| A_{R_k}^{t} \big| = n-k+1$. 
At step $t$ the set $A_{L_k}^{t}$ can be built with at most $k-1$ comparisons.
Note that, for the symmetry of the problem, (i) we can avoid examining $A_{R(k)}^{t}$; (ii) there is no difference in solving for $k>\lceil \frac{n}{2} \rceil $ or  $k<\lceil \frac{n}{2} \rceil $; (iii) finding $k=\lceil \frac{n}{2} \rceil $ is the most demanding case.
We can now express the selection problem in  algorithmic form:

\begin{algorithm}[H]
\SetAlgoLined
\Input{$A=(a_1 \ldots a_n)$ and $k$}
\While{$\big| A_{L_k}^{t} \big| \ne k-1$}{
Select element $a_{k}$ as pivot $a^t_{k}$ \;
Build the  set $A_{L_k}^{t}$, with a series of swaps relying on comparisons with $a^t_{k}$ \;
Evaluate the cardinality of $A_{L_k}^{t}$\;
}
\Output{$a_{k}$ }
\end{algorithm}
%

\noindent
The key parts of the algorithm can be recognized in the code listing (\ref{DIVA:algo}), distilled from  function \texttt{quickselectFS.m}.
The code uses three variables to control the progression of $A_{L_k}^{t}$ and  $A_{R_k}^{t}$: one is \texttt{position} and the others are two `sentinels' \texttt{left} and \texttt{right} such that $\mbox{\texttt{left}} \le \mbox{\texttt{position}} $ and $\mbox{\texttt{right}} > \mbox{\texttt{position}} $ at each iteration step $t$. 
Note that each change of \texttt{position} is associated to a swap operation.
Note also that, for the symmetry of the problem, we could focus only on the left part of the array; this means that the loop terminates when  $ \mbox{\texttt{left}}  = \mbox{\texttt{position}} +1 = k+1$. 
%
A last remark is on the intense \textcolor{blue}{\texttt{for}} cycle at line 13, which runs only until \texttt{right}-1: this is because $A(\mbox{\texttt{right}})$ receives the pivot element at line 10 and thus the \textcolor{blue}{\texttt{if}} statement at line 14 is never true when $i=\mbox{\texttt{right}}$.  

\begin{figure}[b!]
\begin{center}
\begin{minipage}{0.85\textwidth}
\begin{lstlisting}[caption={Code distilled from \texttt{quickselectFS.m}. Also available as R, \proglang{C} and C-mex files. \label{DIVA:algo} } ]
function [kE] = quickselectFS(A,k)
% Finds the k-th order statistic using SimpleSelect
left 		= 1;
right		= numel(A);
position	= -1;
while (position~=k)
    %% Part one: select element in position $k$ as pivot
    pivot    = A(k);					%| 
    A(k)     = A(right); 				%| Swap
    A(right) = pivot;					%| 
    position = left;                    %    c2
    %% Part two: build set $A_{L_k}^{t}$ 
    for i=left:right-1
        if(A(i)<pivot)
            buffer		= A(i);			%| 
            A(i)		= A(position);	%| Swap
            A(position)	= buffer;		%| 
            position	= position+1;   %    c1
        end
    end
    A(right)	= A(position);
    A(position)	= pivot;
    %% Part three: evaluate cardinality of $A_{L_k}^{t}$
    if (position < k)
        left  = position + 1;           %    c2
    else
        right = position - 1;
    end
end
\end{lstlisting}
\end{minipage}
\end{center}
\end{figure}

\section{Counting comparisons}
\label{time-complexity-formal:sec}

This section shows that \textsf{simpleselect} performs like \textsf{quickselect} and is therefore suitable to the applications discussed in Section \ref{robstat:sec}. 
Abandoning recursion precludes the derivation of theoretical time bounds using recurrence equations. We therefore adopt a simple counting approach. 

\subsection{Worst case}
\begin{figure}[t!]
\includegraphics[width=0.499\textwidth,trim=30 80 30 120,clip]{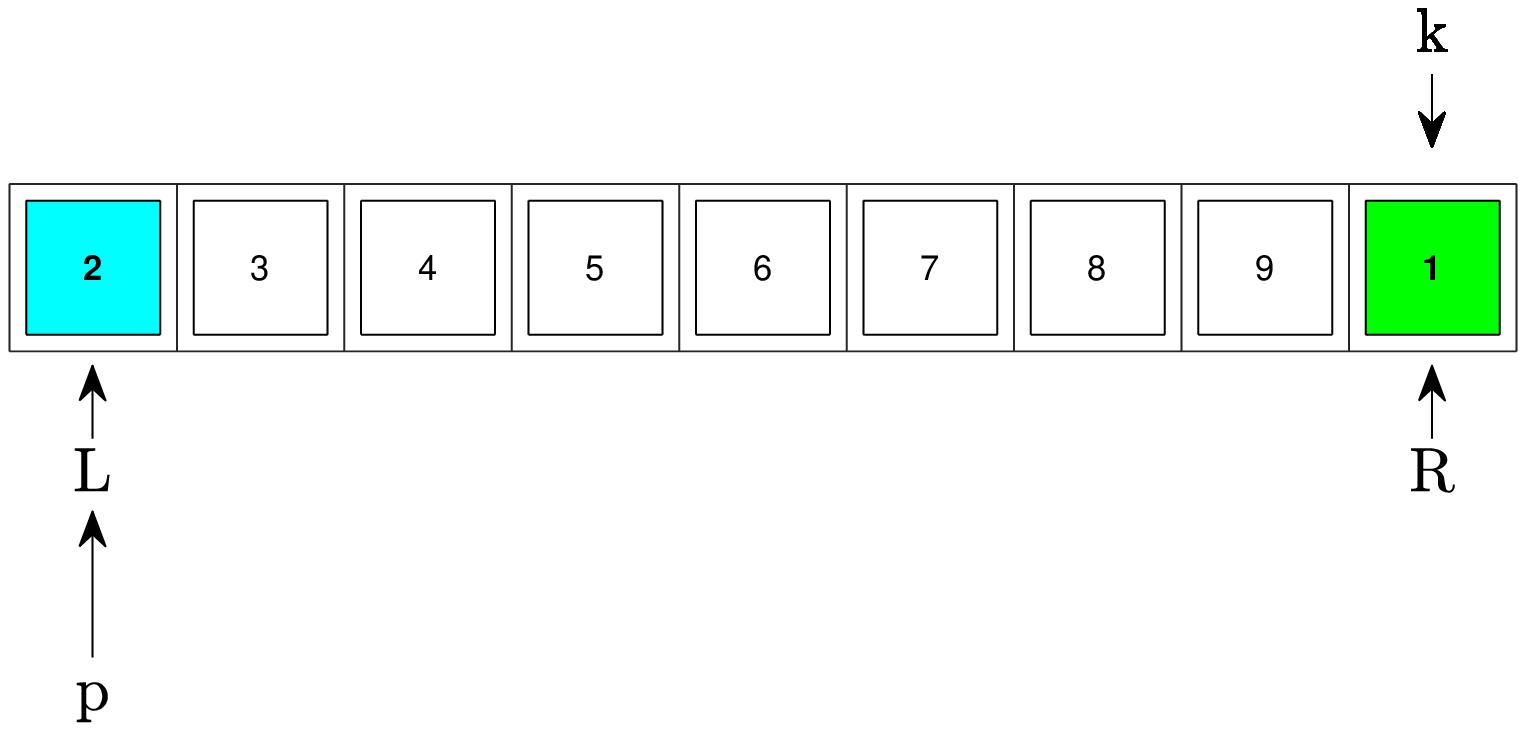}
\includegraphics[width=0.499\textwidth,trim=30 80 30 120,clip]{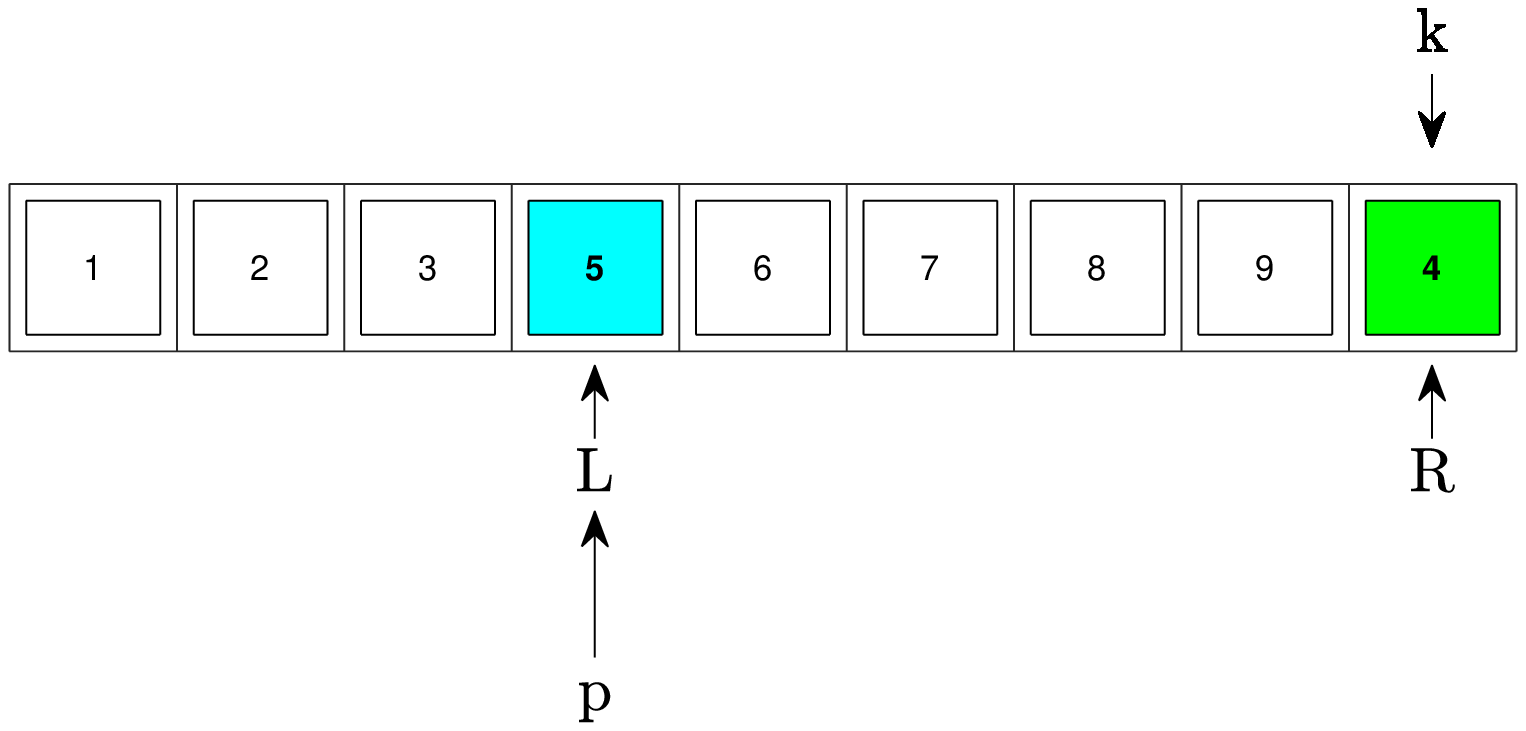}
\caption{\textsf{simpleselect} in the worst case. Left panel, first \textcolor{blue}{\texttt{while}} iteration: $|A_{L_9}^{1}| = 0$. Right panel, forth \textcolor{blue}{\texttt{while}} iteration: $|A_{L_9}^{4}| = 3$. To reproduce, use \mcode{quickselectFS_demo(A,k)} with $A = [2, 3, 4, 5, 6, 7, 8, 9, 1]$ and $k=9$.\label{demo:fig}}
\end{figure}

We start counting the number of comparisons $c(n)$ in the worst case, occurring when at each cycle $t$ the set $A_{L_k}^{t}$ contains exactly $t-1$ elements.
This happens when we look for the maximum ($k=n$) among elements in increasing order except the last containing the minimum ($a_{1} < a_{2} < \ldots < a_{n-1}$ and $a_{n} = a_{(1)}$).
In this case (and the symmetric one for $k=1$), variable \texttt{position} is never modified inside the \textcolor{blue}{\texttt{for}} cycle.
Figure \ref{demo:fig}, produced with a function written to demonstrate the dynamic of \textsf{simpleselect}, illustrates the status at the first and forth \textcolor{blue}{\texttt{while}} iteration of an array with this unfortunate order.  

As comparisons are done at lines $6$, $14$ and $24$ of the code listing (\ref{DIVA:algo}), we specify the count breakdown with $c(n) = c_{6}(n) + c_{14}(n) + c_{24}(n)$.  
Then we indicate with $(t,l,r,p)$ the status  of variables \texttt{left}, \texttt{right} and \texttt{position} at step $t$. With this notation we have:
\begin{eqnarray}
(t,l,r,p)=(1,1,n,1)  & \Rightarrow  & c(n) = 1+(n-0)+1   \nonumber  \\
(t,l,r,p)=(2,2,n,2)  & \Rightarrow  & c(n) = 1+(n-1)+1   \nonumber  \\
%
%
                                                                 & \vdots   & \nonumber \\
(t,l,r,p)=(n,n,n,n) &  \Rightarrow  & c(n) = 1+(n-(n-1))+1   \nonumber 
 \end{eqnarray} 
Therefore, the worst number of comparisons is quadratic:
\begin{equation}
c(n) = 2n + n^2 - \sum_{i=0}^{n-1} i = 2n + n^2 - \frac{(n-1)n}{2}  = \frac{n^2 + 5n}{2}.
\label{worstcase:eq}
\end{equation}
%
%
%
\begin{figure}[b!]
\centering
\begin{minipage}{0.49\textwidth}
\includegraphics[width=1.1\textwidth]{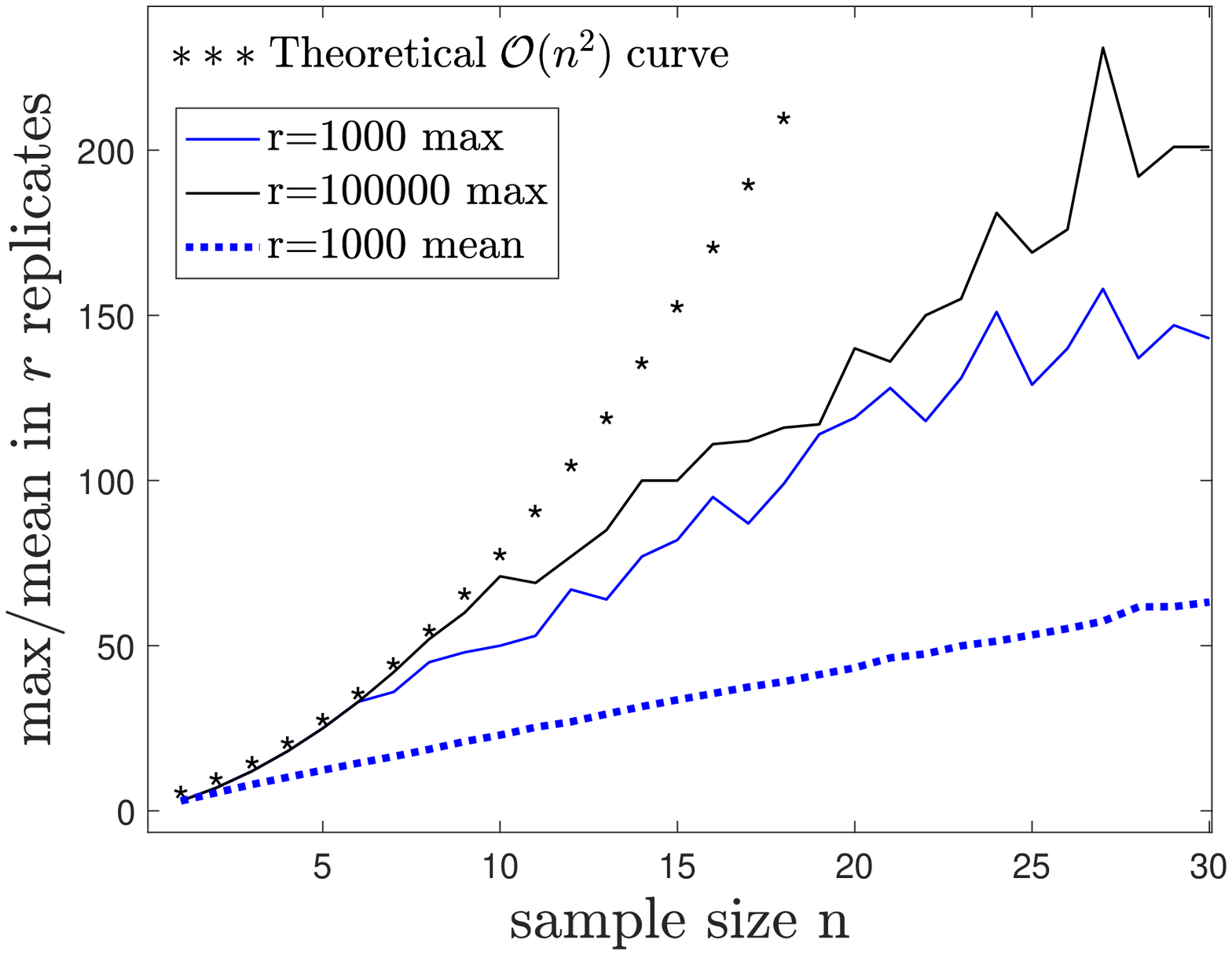}
\end{minipage}
\hfill
\begin{minipage}{0.49\textwidth}
\includegraphics[width=1.1\textwidth]{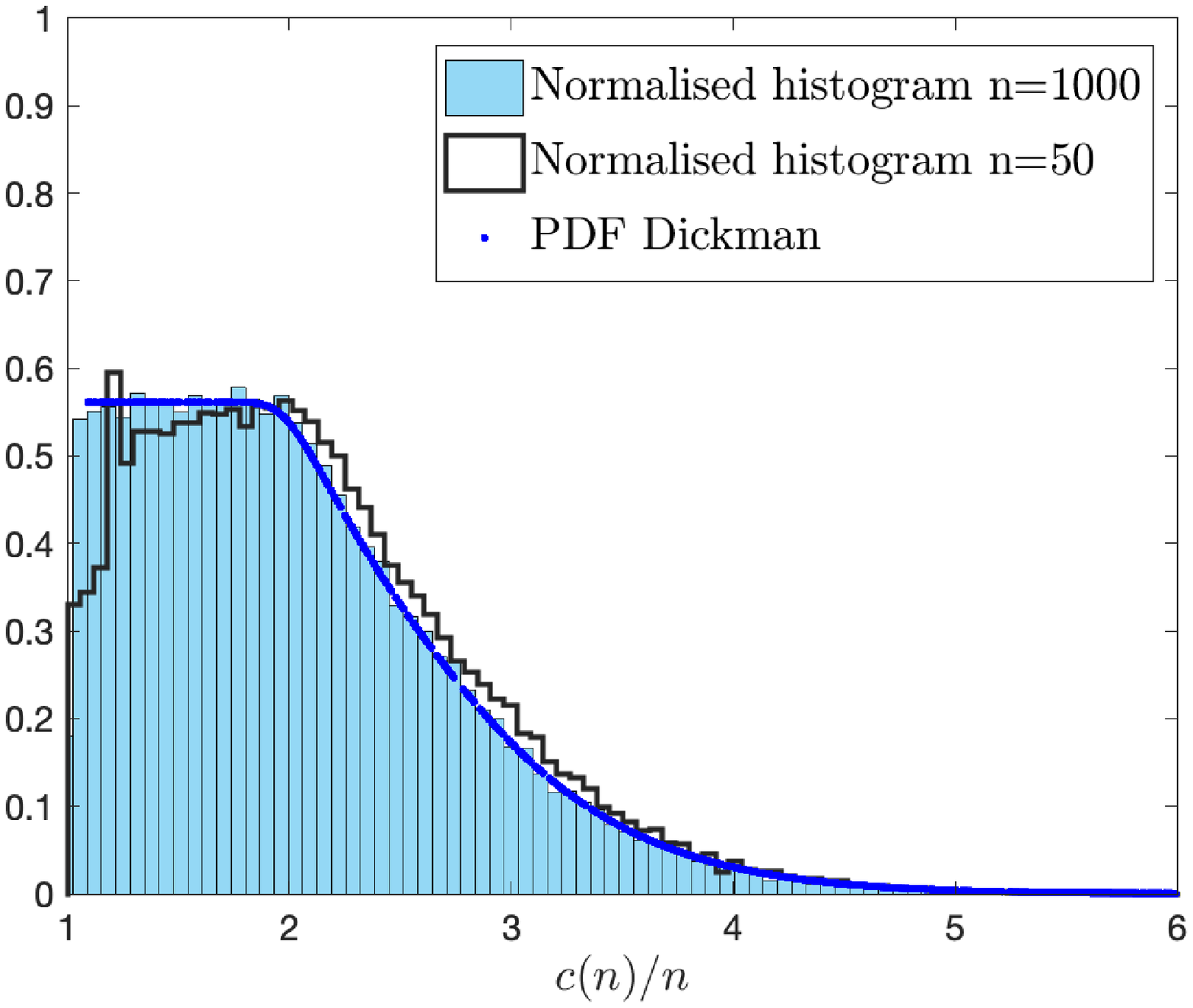}
\end{minipage}
\caption{\label{fig:comparisons_findmax} \textsf{simpleselect}. Empirical number of comparisons $c(n)$ for finding the maximum in a uniformly generated sample. 
Left panel: $\max(c(n))$ and $\mbox{mean}(c(n))$ in $r$ replicates; the line fit on the mean case is $\hat{c}(n)=2\,n$. 
Right panel: empirical distribution of $c(n)/n$ for $n=1000$ (bar style) and $n=50$ (stairs style), with superimposed the Dickman distribution (translated by 1) obtained using FSDA function \texttt{vervaatxdf} with parameter $\beta=1$.  
%
}
\end{figure}
\noindent 
Note that $c_{6}(n)$ and $c_{24}(n)$ only involve \textit{index comparisons} with short integers, which typically cost much less than a \textit{data comparison} $c_{14}(n)$ on the array content. If we ignore them, equation~(\ref{worstcase:eq}) reduces to $ c(n) = (n^2 + n)/2 $.

\subsection{Average case}
We can reason about how the linear complexity is achieved in practice with average case considerations, which the simplified code makes almost trivial.
At the first execution of the \textcolor{blue}{\texttt{while}} statement, a comparison is done at line 6 to check the exit condition, then one is done at line 24, and another $n$ are executed inside the  \textcolor{blue}{\texttt{for}} cycle of line 13-20. 
The exit condition set on \texttt{position}  is approached:
\begin{itemize}
\item[c1:] at line 18, where \texttt{position} is incremented by a step whenever  $ a_i < \mbox{\texttt{pivot}}$;
\item[c2:] at line 25, where 
\texttt{left} jumps to the cell after \texttt{position} and this makes also \texttt{position} to make a step ahead, being set to \texttt{left}  at line 11.
\end{itemize}
If $A$ is a random sample of elements extracted from the same distribution, we expect $(n-1)/2$ increments of \texttt{position} in the initial scan of $A$ (see Annex \ref{result1:sec}).
Likewise, the increment of variable \texttt{left} at lines $25$ and $11$ - and similarly for variable \texttt{right} - which further reduces the distance to \texttt{position}, is done at most once.
Then, at the second execution of the \textcolor{blue}{\texttt{while}} statement the \textcolor{blue}{\texttt{for}} cycle is expected to run on about (less than) $n/2$ array cells (actually, $(n-1)/2-1=(n-3)/2$). And so on for the subsequent steps: 
\[
\begin{array}{lcll}
(t,l,r,p) = (1,1,n,1)  & \Rightarrow  &  c(n) = 1+n+1  \nonumber  \\

(t,l,r,p) = (2,\cdot,n,\cdot) & \Rightarrow  &   c(n) = 1+n/2+1 \nonumber  \\
(t,l,r,p) = (3,\cdot,n,\cdot) & \Rightarrow  & c(n) = 1+n/4+1 \nonumber  \\
                                                                & \vdots   & \nonumber \\
(t,l,r,p) = (\log_2 n,n,n-1,n) &  \Rightarrow  &  c(n) = 1+n/2^{(log_2 n -1)}+1 \nonumber  
 \end{array} 
\]
The total therefore is:
\begin{align}
c(n) & =  \displaystyle 2 \log_{2} n  +  n \sum_{t=1}^{\log_{2} n} \frac{1}{2^{t-1}} & = &\; 2 \log_{2} n   +  n \sum_{i=0}^{\log_{2} n -1} \frac{1}{2^{i}}  &  \nonumber \\
        & =   \displaystyle 2 \log_{2} n  +  n + n \sum_{i=1}^{\log_{2} n - 1} \frac{1}{2^{i}} & =  &\; 2 \log_{2} n   + n + n (1 - \frac{1}{2^{\log_{2} n - 1}})  &   \label{averagecase1:eq}  \\
         & =   \displaystyle 2 \log_{2} n  +  n + n (1 - \frac{2}{2^{\log_{2} n}})  & = &\; 2 \log_{2} n   +   n + n (1 - \frac{2}{n})  &  \nonumber \\
         & =   \displaystyle 2 \log_{2} n + 2 n - 2 & < & \;   2 n + 18 \qquad \mbox{for} \; n<1000 &  \label{averagecase2:eq} 
\end{align}
where the partial sum of the first $\log_2 n - 1$ terms of the geometric series in (\ref{averagecase1:eq}) is computed with standard algebraic steps, and the final upper bound (\ref{averagecase2:eq}) reduces the logarithmic term to a constant that increases barely with $n$ compared to $2n$.
%
Note that the conclusion is in line with the empirical results of Figure \ref{fig:comparisons_findmax}, which indeed bode for a $2n$ term, and with known asymptotic results based on the recurrent relations of recursive partitioning methods \citep[as in][Theorem 1]{Mahmoud_et_al:1995}. 

\subsection{Number and empirical distribution of comparisons}
\label{dickman:sec}

The left panel of Figure~\ref{fig:comparisons_findmax} shows  with symbol `$*$' the progression of equation~(\ref{worstcase:eq}) and the actual number of comparisons required by \textsf{simpleselect} for finding in $r$ replicates the maximum in a set of $n$ integers extracted uniformly between 1 and (to avoid repetitions) $10000\, n$. 
For small sample sizes the two curves of the maximum follow the quadratic shape of equation (\ref{worstcase:eq}), as the $r$ replicates are enough to fall in the worst possible scenario (extract the $n$ values in the above mentioned order). Then they start growing linearly, with an \textit{empirical worst-case} of approximately $5\,n$ (the fit is for $r=100000$ and $n$ up to $1000$).
%
Similarly, the third (dotted) line shows that \textit{on average}  $2\,n$ comparisons are sufficient  to find the maximum. 
%
%
%


We also checked the empirical distribution of $c(n)/n$: for sufficiently large $n$ it follows the Dickman distribution shown in the right panel of Figure \ref{fig:comparisons_findmax}, which is the limiting distribution for  Hoare's \texttt{Find} (see \cite{hwang_tsai:2002} and related works by \cite{Giuliano_etal:2018,Goldstein:2018}). 
It is remarkable that the Dickman adaptation is rather good also for small sample sizes ($n=50$). 
Given the centrality of the Dickman distribution in this context, we provide functions for the computation of the pdf, cdf and the generation of random variates of the wider Vervaat class, following the works mentioned in Appendix \ref{vervaat:sec}. 
These functions can be used to study partitioning algorithms different from Hoare's \texttt{Find}, which can lead to other limiting results \citep{Mahmoud:2010}. 

%

\subsection{Run-time results}
\label{time-complexity-empirical:sec}

Our baseline for the run-time results is the internal (built-in) \texttt{sort} function -- $SORT_{int}$ -- which a typical \proglang{MATLAB} user would use because the standard distribution do not have functions dedicated to order statistics (\texttt{prctile} and \texttt{median} use \texttt{sort}).  
For assessing the actual performances of \textsf{simpleselect}, we consider two execution modes of our implementation: the just-in-time (\texttt{jit}) compilation where \proglang{MATLAB} directly analyses and translates the source code \textit{during} a run (this is the standard execution modality of \proglang{MATLAB}, also available in R), and the \texttt{mex}-file mode where we compile a \proglang{C}-code instance of \textsf{simpleselect} into machine-code \textit{before} any subsequent execution. 
The two instances, identified later with $SS_{jit}$ and $SS_{mex}$, are then compared with the classic implementation of \textsf{quickselect} of Numerical Recipies in \proglang{C} \citep[][Section 8.5]{NRC:1992} -- $NURE_{mex}$ -- and the \textsf{introselect} available in the \proglang{C++} function \texttt{nth\_element} --  $NthEL_{mex}$ -- both compiled as \texttt{mex}-file.
The latter is also distributed as \texttt{mex}  file by \cite{nthelement:2013}, but uses undocumented calls that may change and break in the future.

%
\begin{algorithm}[tb!]
\vspace{2mm}
\SetAlgoLined
\Input{$pdfName$, $nSet$, $kSet$, $r$}
\For{n=nSet}{
\For{k=kSet}{
\For{i=1:r }{
	$A_{n} \leftarrow$  $n$-sample from distribution $pdfName$\;
	$\mbox{\texttt{EtimeCount}}(n,k,i) \leftarrow$ \texttt{tic SimpleSelect($A_{n},k$) toc}\;
	$\mbox{\texttt{NumOpCount}}(n,k,i) \leftarrow$ \texttt{SimpleSelectOpCount($A_{n},k$)}\;
}
}
}
$\mbox{\texttt{grpVar}} \leftarrow [nSet , kSet]$\; 
$\mbox{\texttt{whichStat}} \leftarrow [\mbox{`median' , `mean' , `max' , `min'}]$\;  
\Output{$\mbox{\texttt{NumOpStats}} \leftarrow \mbox{\texttt{grpstats(NumOpCount,grpVar,whichStat)}} $ \; \\
$\mbox{\texttt{EtimeStats}} \leftarrow \mbox{\texttt{grpstats(EtimeCount,grpVar,`median')}} $ \;}
\vspace{5mm}
\caption{\label{simcode:fig} Pseudo-code used to assess execution time and number of basic operations of \textsf{simpleselect}. The integer array $kSet$ contains the order statistics of interest. Random samples are generated from the distribution specified by $pdfName$ with sizes defined by the integer array $nSet$.  Performances in output are computed as a function of $nSet$ and/or $kSet$ on the basis of $r$ replicates.}
\end{algorithm}


\begin{figure}[t!]
\begin{minipage}{0.49\textwidth}
\includegraphics[width=\textwidth]{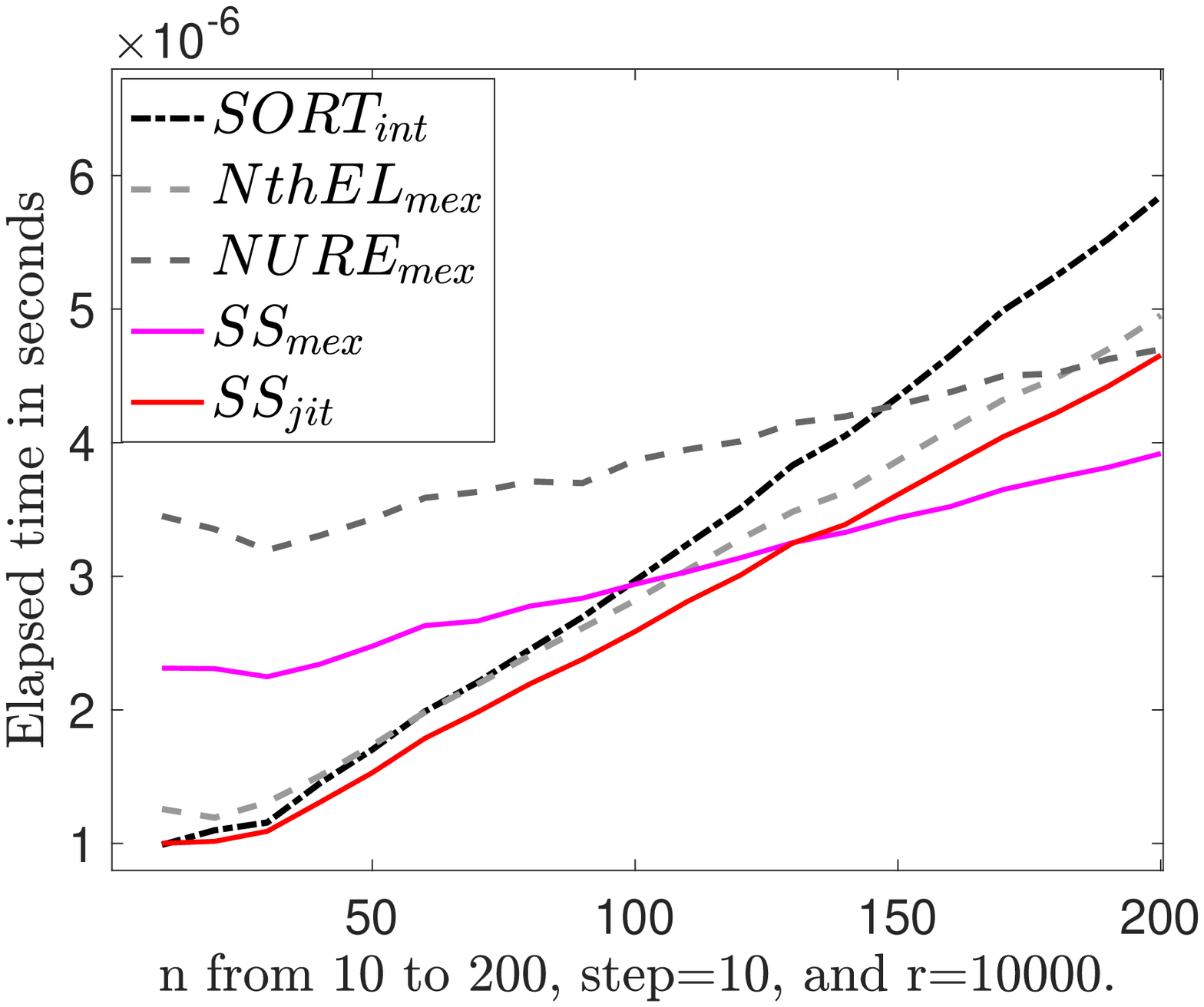} 
\end{minipage}
\begin{minipage}{0.49\textwidth}
\includegraphics[width=\textwidth]{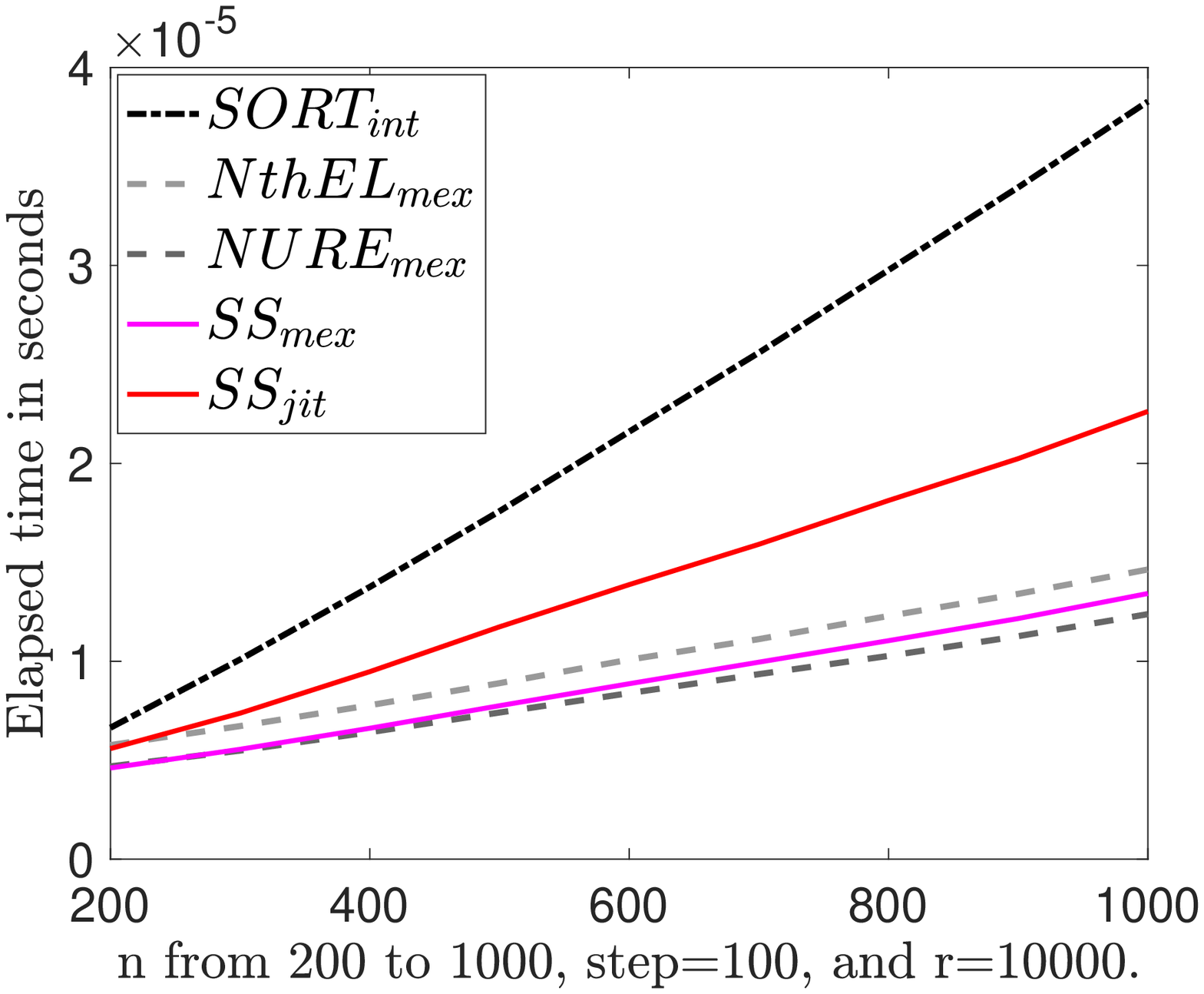} 
\end{minipage}
\\[5mm]
\begin{minipage}{0.49\textwidth}
\includegraphics[width=\textwidth]{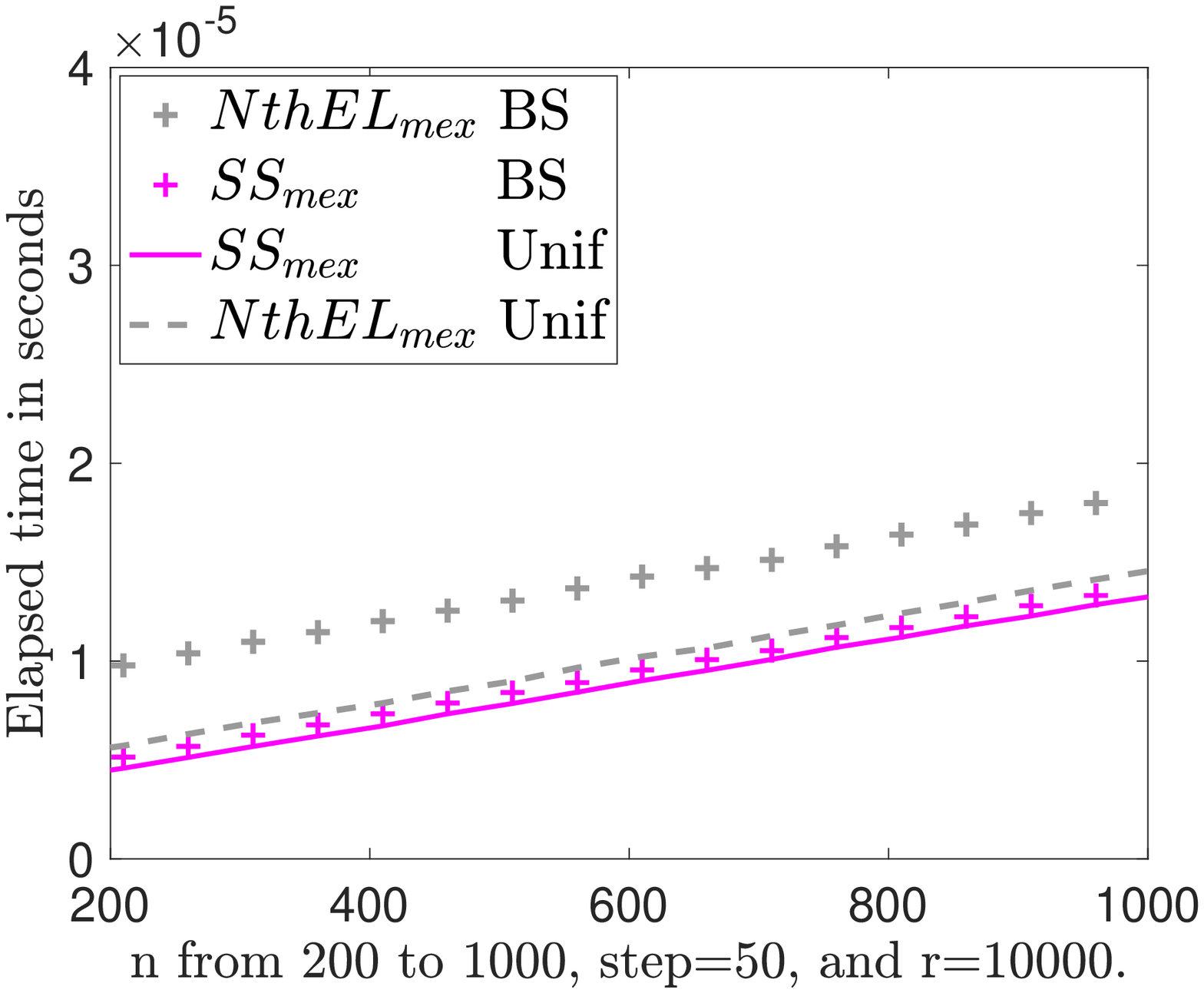} 
\end{minipage}
\begin{minipage}{0.49\textwidth}
\includegraphics[width=\textwidth]{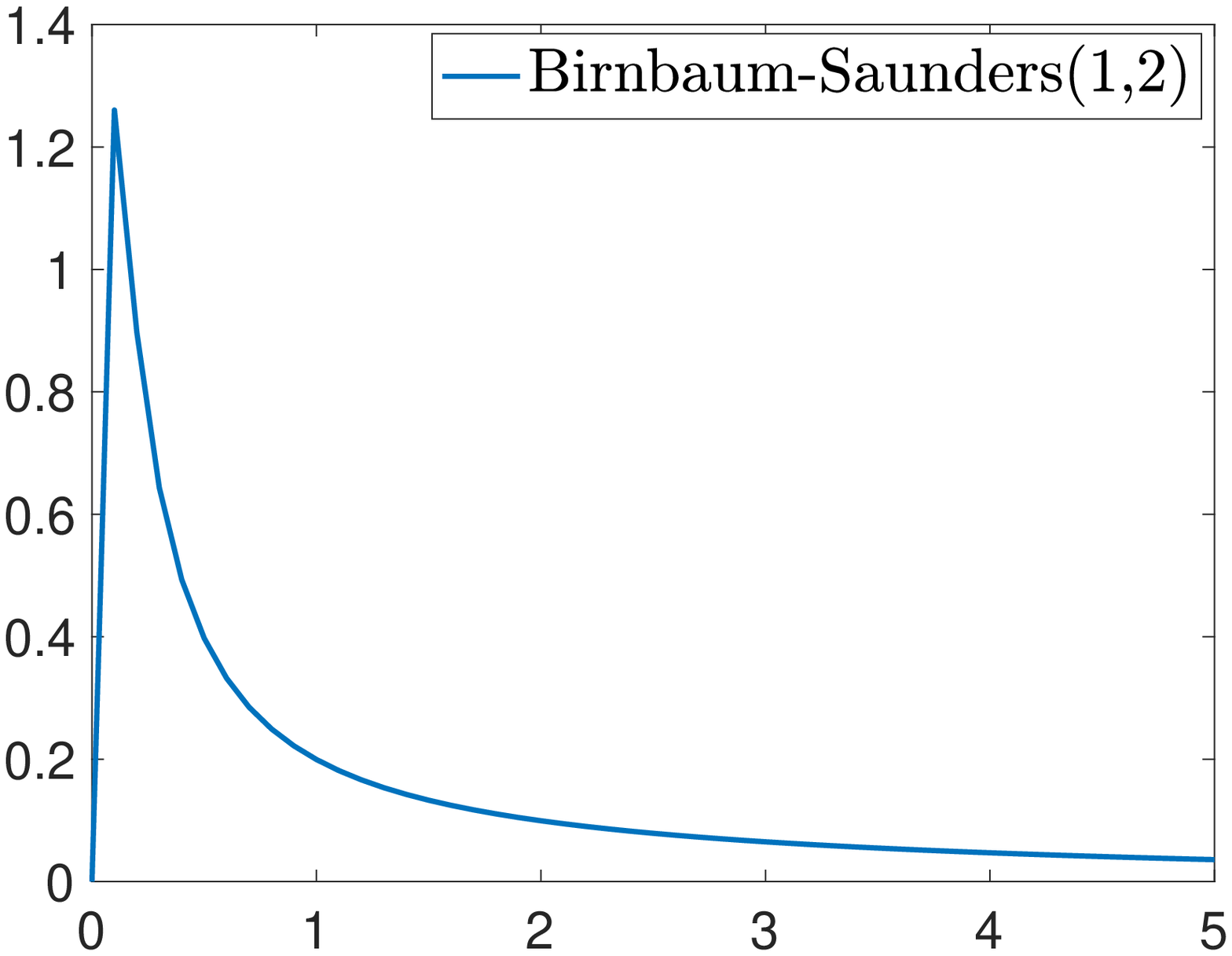} 
\end{minipage}
\caption{\label{etime_comparison:fig} Finding the median. Top panels: elapsed time 
in seconds of \textsf{simpleselect} ($SS$),  \textsf{quickeselect} as in Numerical Recipies in \proglang{C} ($NURE$), \textsf{introselect} as in the C++  \texttt{nth\_element} function ($NthEL$) and the \proglang{MATLAB}-internal \texttt{sort} ($SORT$). Bottom-left panel: the `$+$' symbols in gray and magenta refer respectively to \textsf{introselect} and \textsf{simpleselect} when $A$ is extracted from the  Birnbaum-Saunders (\texttt{BS}) distribution on the bottom-right; the other two lines are obtained with the uniform distribution (\texttt{Unif}), as in the top panels.}
\end{figure}


The assessment is structured along the pseudo-code  \ref{simcode:fig} (the actual code is available as supplementary material).
The time results obtained for finding the lower median for $n$-values ranging between 10 and 1000, are reported in Figure \ref{etime_comparison:fig} as median time over 10000 replicates. Therefore, the possibility of introducing a bias such as the potential initial latency of the \texttt{jit} compilation, is completely removed. 
%
The results show that:
\begin{enumerate}
\item 
Top panels. The plain \proglang{MATLAB} implementation of \textsf{simpleselect} ($SS_{jit}$) is  competitive even for small sample sizes ($n<100$), but replacing the \texttt{sort} function provides limited time drop. 
The advantage increases considerably for larger sample sizes and becomes neat if the \texttt{mex}-compiled version is used; note also that its performance is in line with that of $NURE_{mex}$ and $NthEL_{mex}$. 
\item 
Bottom panels. A key advantage of \textsf{simpleselct}, and comparable algorithms relying only on the pairwise order of the elements in $A$,  is that performance is independent from the distribution of the data. On the contrary   \textsf{introselect}, which tries to reach optimal worst-case performance by exploiting statistics (means and medians) on partitions of the data, may suffer from peaked data distributions like the one in the figure. 
\cite{Tibshirani:08} contains an extensive assessment exercise for a similar algorithm, the \textsf{binmedian}, which is indicative of the complications linked to the adoption of sophisticated solutions in real applications.
\end{enumerate}



%

\section{Use of selection in robust statistics}
\label{robstat:sec}

Many robust methods require order statistics to identify an outlier-free subset in a set of $p$-variate observations $X_n =  \{x_1 , \ldots , x_n\}$. 
We illustrate with two case studies the advantages of adopting \textsf{simpleselect} for this purpose: in the first situation the required order statistic remains fixed at a same point (typically the median), while in the other it increases from $p+1$ to $n$ leaving, from a certain progression point, most of the largest values on the right side of the array.  

\begin{description}

\item[Minimum Covariance Determinant (MCD).]


The MCD tries to identify the subset of $h$ out of $n$ $p$-variate observations giving rise to the smallest determinant of the covariance matrix \citep{Rou:84,Rou:85}. 
The exact solution requires to evaluate ${n \choose h}$ cases, which in general is hard to compute being $(\frac{n}{k})^{k} \le {n \choose h} \le (\frac{en}{k})^{k}$ \citep[][p. 1186]{Cormen:2009}. 
Fortunately the MCD has an approximate solution \citep{RouVDri:99} that relies on taking at random many initial subsets and applying on each of them a fixed point iteration scheme with this property: if at step $t$ we have an $h$-subset $H_{t}$ with empirical mean and covariance matrix $\hat{\mu}_{t}$ and $\hat{\Sigma}_{t}$, and the (Mahalanobis) distance of observation $i$ is $d_{i} = d(x_{i}, \hat{\mu}_{t} , \hat{\Sigma}_{t})$, then the new $h$-subset $H_{t+1}$ formed by the observations with squared distances $d^{2}_{i} \le d^{2}_{(h)}$ is such that $| \hat{\Sigma}_{t+1} | \le | \hat{\Sigma}_{t}|$. 
Therefore, each iteration requires the computation of the order statistic $d_{(h)}$. 
Typically $h$ is set to $[(n + p + 1)/2]$, which is around the median, and is possibly increased with a weighting step to improve the estimator's efficiency. 
The initial random subset $H_0$ is of size $p+1$, which reduces the chance to embed outliers in computing the initial estimates $\hat{\mu}_{0}$ and $\hat{\Sigma}_{0}$. 
The loop continues until the equality condition on the determinant of the two covariance matrices is satisfied. Usually few iterations are sufficient to reach convergence, but the number of subsets to sample and iterate can be in the order of some thousands, and so are the applications of the appropriate order statistic computation. 

\item[Forward Search (FS)]

The FS \citep[][]{arc:2004}  adapts the value of $h$ to the data with an iteration procedure combined with a testing step. The iteration generates a sequence of parameter estimates, while the testing determines the $h$ value and detects the outliers on the basis of such parameters.
More precisely, the iteration starts from a very robust fit to a few carefully selected observations, say $m_0 = p+1$. 
Then it takes the $m_1 = m_0+1$ observations with the smallest squared  distances from the robust fit of $\hat{\mu}_{0}$ and $\hat{\Sigma}_{0}$. 
At step $m_1$  the parameter estimates are again computed and the process is iterated until all units are included ($m=n$).
Therefore, each step of the iteration requires (i) a selection application to determine the $d_{(m+1)}^{2}$ smallest order statistic in the set of $i=1,\ldots,n$ distances computed on the basis of $\hat{\mu}_{m}$ and $\hat{\Sigma}_{m}$ and (ii) $n$ comparisons to identify the observations $i$ for which $d^{2}_{i} \le d^{2}_{(m+1)}$. Therefore, there are $n - m_0 + 1$ steps requiring the computation of different (increasing) order statistics.

\end{description}

We illustrate the use of \textsf{simpleselect} in MCD and FS focusing on the potential benefit of its optional ``oracle'' parameter, which provides the index $j$ of an element in $A$ that might contain the desired $k$-th order statistic or be close to it. This option simply swaps $A(j)$ with $A(k)$ before starting the process; unless badly chosen, the initial guess on the pivot reduces the chance of falling into the worst case and improves the average case performance. 
For example, if at a certain FS step $m$ the variable \texttt{minMDindex} contains the index of the minimum of Mahalanobis distance among the units which form the group of potential outliers, then to increase by one unit the dimension of the basic subset \texttt{bsb} one could use option at line $9$ below where $j \gets \mbox{\texttt{minMDindex}}$, instead of the standard \textsf{simpleselect} without oracle (line $6$) or the \texttt{sort} that we use as a baseline (line $3$): 
\vspace{-4mm}\begin{lstlisting}[frame=none]
            switch bsb_update_by
                case 'sort'
                    [~ , zsi] = sort(MD);
                    bsb = zsi(1:m+1);
                case 'quickselectFS'
                    k=quickselectFS(MD,m+1);
                    bsb=(MD<=k);
                case 'quickselectFS_oracular'
                    k=quickselectFS(MD,m+1,minMDindex);
                    bsb=(MD<=k);
             end
\end{lstlisting}

In the MCD a similar approach is used in the iterative re-weighted least squares step, where the location and shape matrix are updated repeatedly till convergence. In this case we need to find the subset of $h$ observations with smallest covariance determinant and there is no counterpart to line $9$; the best we can try is to choose randomly a candidate between the units that are not in the current set of $h$ observations, as in line $12$ here:  
%
\vspace{-4mm}\begin{lstlisting}[frame=none , firstnumber=12] 
	k = quickselectFS(MD,h+1,h+1+randi(n-h-1,1,1));
\end{lstlisting}

\begin{figure}[t!]
\includegraphics[width=0.45\textwidth]{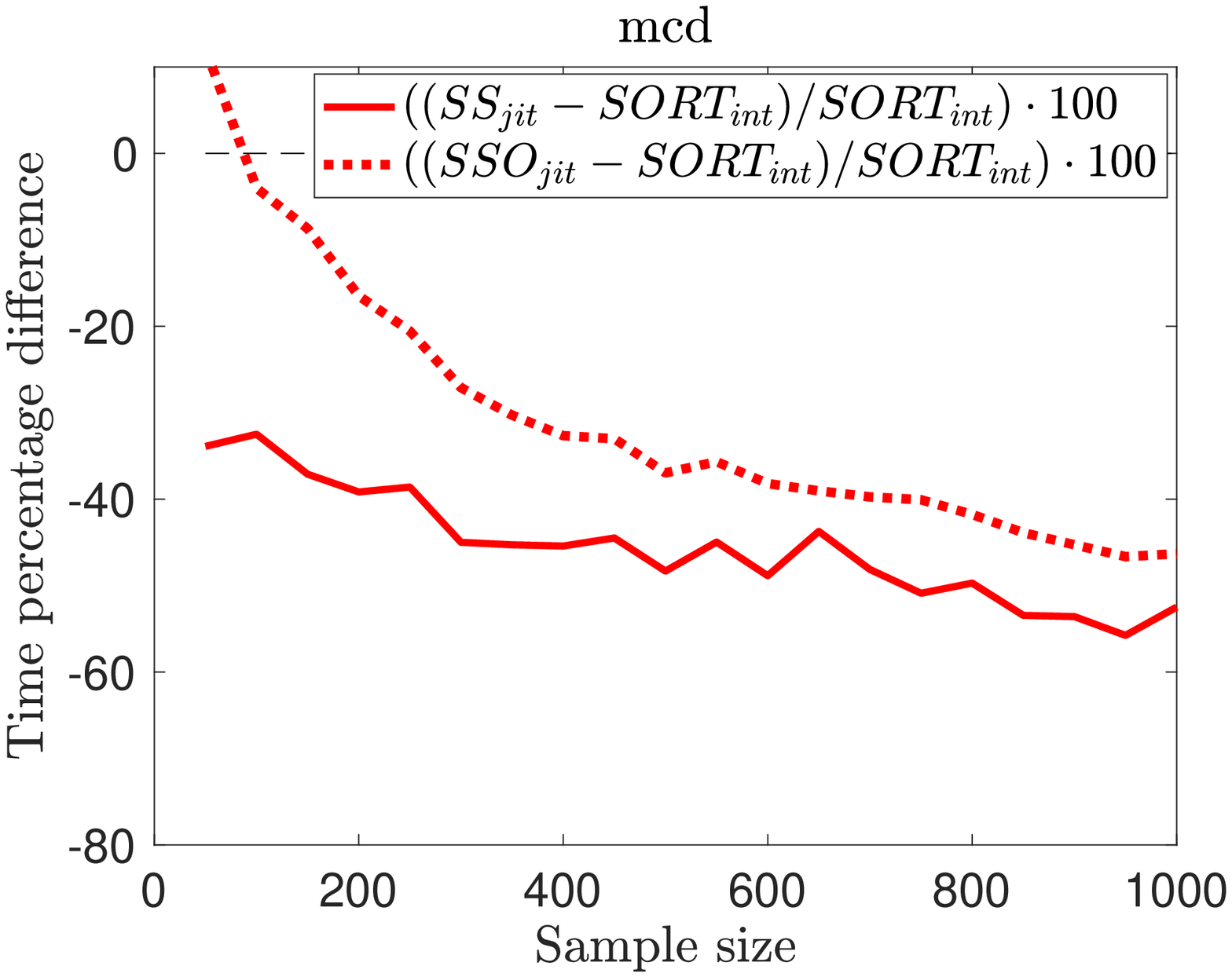} 
\includegraphics[width=0.45\textwidth]{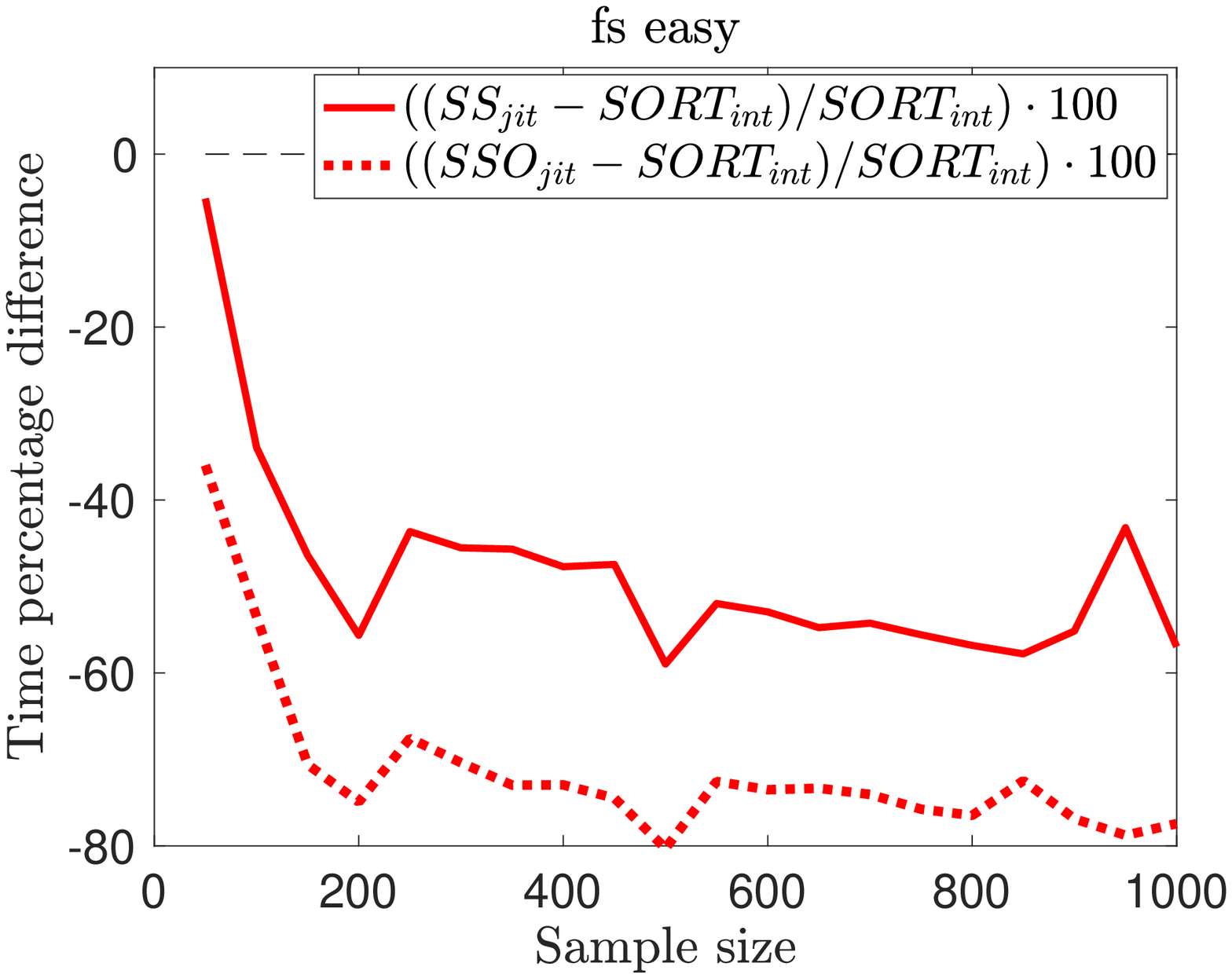} \\[3mm]
\includegraphics[width=0.45\textwidth]{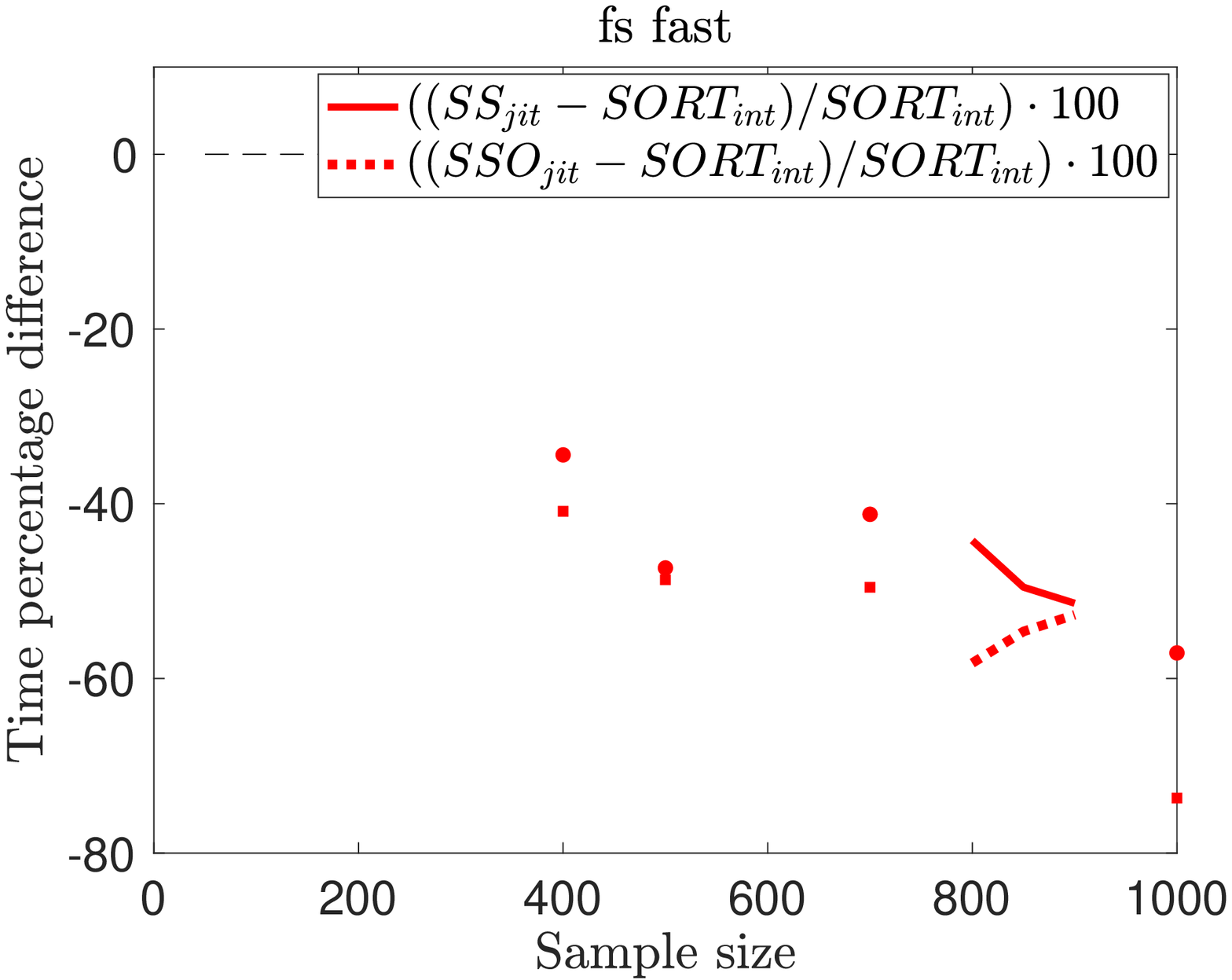} 
\includegraphics[width=0.45\textwidth]{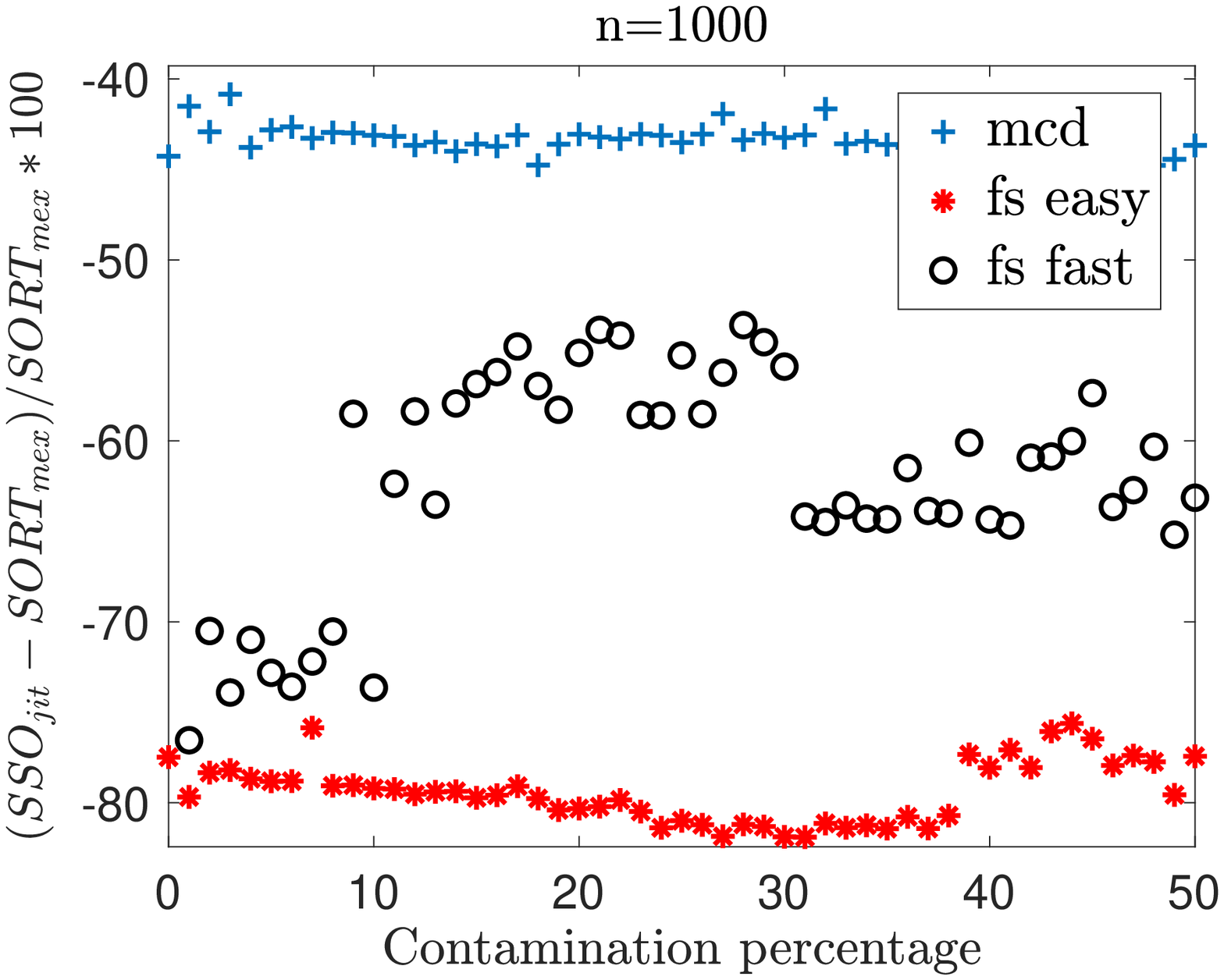} 
%
\caption{\label{SSO:fig} Effect of the oracular option of \textsf{simpleselect} in the MCD and two FS implementations, on data generated from a bi-variate normal distribution. Top and bottom-left panels: time percentage reduction w.r.t the \texttt{SORT} baseline with and without oracular option, for various sample sizes. Bottom-right panel: time reduction for fixed $n=1000$ and varying contamination level. }
\end{figure}

The top-left panel of Figure \ref{SSO:fig} shows that the random oracular choice does not help the MCD: in fact, the \texttt{SSO} curve is well above \texttt{SS}. For the FS the situation is reversed and the advantage of \texttt{SSO}  is neat, but we have to distinguish between two implementations of the algorithm: one follows step by step the original FS formulation \citep{a+r:2000,arc:2004} in the \proglang{R} package \texttt{forward} and the more recent FSDA function \texttt{FSMmmdeasy.m}; the other is a fast (but hardly readable) version of the algorithm \citep{RiPeCe:15} that in most of the forward steps updates the subset with logical operations instead of using \texttt{sort} or \texttt{quickselectFS}. The fast version -- FSDA function \texttt{FSMmmd.m} -- limits the application of \texttt{quickselectFS} to steps where more than one unit exit from the subset, the so called ``interchange''. The effect is visible in the bottom-left panel of  Figure \ref{SSO:fig}, where the time information is available only episodically and for the larger $n$ values. However, the indication is that even in presence of interchange, and therefore uncertainty in the choice of $j$, the oracular option speeds up the method. 

Given that the interchange increases if data are contaminated, we study the effect of creating two separate groups by adding a fixed shift to an increasing number of units in the normal bi-variate sample passed to MCD and the two FS implementations. 
As expected, the bottom-right panel of  Figure \ref{SSO:fig} shows that the time gain in the MCD remains stable (towards $-40\%$), being the algorithm not subject to interchange by definition. 
In the fast FS the effect is strongest for small (up to $10\%$) contamination percentages and returns visible for large ones (above $35\%$); this is due to the fact that we initiate the FS with contaminated units, therefore initially we expect to fall in very unstable estimates and strong interchanges.  
In the standard FS the gain is remarkable (about $-80\%$) and  increases as the contamination percentage approaches $40\%$; here the share due to the contamination  is more difficult to appreciate, as most of the calls to \texttt{quickselectFS} do not depend on the interchange.   



\section{Extension to weighted selection}
\label{weighted:sec}

%

A number of statistical problems reduce to the selection of an order statistic on elements that are assigned with non-negative weights. 
For integer weights, this means choosing the order statistic from an increased set where each element is replicated to the number of the corresponding weight. 
We introduce the general problem focusing on the  \textit{weighted median} (Section \ref{weightedmedian}), which is the 50\% \textit{weighted percentile}.
We implement the weighted percentile as a natural extension of the \textsf{simpleselect} algorithm, in function \texttt{quickselctFSw} (Section \ref{weightedSimpleSelect}). 
We illustrate its application to a computation-intense robust measure of skewness, the medcouple (Section \ref{medcouple:sec}).

\subsection{Weighted median roots and  applications}
\label{weightedmedian}

Finding the median $\tilde{A}$ defined in Section \ref{intro:sec}  also solves the optimization problem  
\begin{equation} 
\tilde{A} = \argmin_a \sum_{i=1}^{n}  |a_i - a| 
\label{weighted_mean:eq}
\end{equation}
Intuitively, the proof relies on the fact that the derivative with respect to $a$ of the sum of the absolute deviations is \(\sum_{i=1}^{n}{ \signum (a_i -a) }   \),
which is zero only when the number of positive terms equals the number of the negative ones, which happens when $a$ is the median.  
If the absolute deviations are weighted by positive quantities $w_i$, then the minimization brings to the weighted median
\begin{equation} 
\tilde{A}_w = \argmin_a \sum_{i=1}^{n}  w_{i}|a_i - a| 
\label{weighted_median:eq}
\end{equation}
This optimization problem has fascinating historical roots  in the \textit{least absolute deviation regression} \citep[][]{Stigler:1984, Farebrother:1990}, formulated already in 1760 by Boscovich and Simpson as a line minimizing the sum of the deviations of the observations from the line. 
%
%
Laplace, in his ``Methode de Situation'' (1818), indicated a solution for the line's slope, finding that the weighted median solves the constrained least absolute deviation regression obtained by replacing in (\ref{weighted_median:eq}) $w_i = | x_i - \overline{x} |$ and   
$a_i = (y_i - \bar{y}) / (x_i - \bar{x})$, being $(x_i , y_i)$, $i=1,\ldots,n$, a two-dimensional sample of points with one independent and one dependent variables respectively. 
Laplace understood that  $\tilde{A}_w$  in (\ref{weighted_median:eq}) is equal to $a_{(k^*)}$, with $k^*$ found by considering the order statistics of the weights $w_{(i)}$ and returning the smallest $k$ associated with the weight whose running sum crosses $50\%$ of the total weight, that is: 
\begin{equation} 
 k^* = \min \left \{ k \mathrel{} \middle| \mathrel{} \sum_{i=1}^{k}  w_{(i)}  \ge \frac{1}{2} \sum_{i=1}^{n}  w_{i}  \right \} 
 \label{kstar:eq}
 \end{equation}
The problem found consolidation with the ``double median'' of \citep{Edgeworth:1888} (the solution for the intercept) and a century later with the simplex algorithm \citep{BarrodaleRoberts:1973,BloomfieldSteiger:1980}.
Modern concepts that revisit and extend these ideas are the  ``dual plot'' and ``regression depth" by \cite{RouHub:99} and the robust time series smoothing and filtering by \cite{DaviesFriedGather:2004,FriedEinbeckGather:2007}.

The weighted median and percentiles find intriguing test cases also in engineering and machine learning, where disposing of efficient algorithms is essential for large-scale applications.
For example, non-linear digital filtering devices embed weighted percentiles for noise cancellation  \citep{AstolaKuosmanen:1997,LinYangGabboujNeuvo:1996}, while medical experiments use them when the precision of individual estimates varies considerably \citep{BDGHB:2016}.
In machine learning it is interesting the case of boosting procedures, which aim generating an accurate prediction by combining several weaker estimators (statistics treats the case in the additive models theory \citep{FriedmanHastieTibshirani:2000}).
%
The original formulation of boosting by  \cite{FreundSchapire:1997} proposes to build the final prediction as weighted average of the individual models, but also shows that the optimal prediction in regression (\texttt{AdaBoost.R2}) should be based on the weighted median of the weak learners. 
Additional motivations for the weighted median were given by \cite{Kegl:2003,Kegl:2004} in view to achieve a certain degree of robustness and by \cite{BertoniCampadelliParodi:1997} for responses in $[0,1]$. 
Recently, \cite{AMR:2022} have shown an application of that approach to sensitivity analysis, where predictions (and therefore  weighted medians) have to be computed a great number of times. 

%


\subsection{Weighted simpleselect}
\label{weightedSimpleSelect}

\begin{figure}[t!]
\includegraphics[width=0.48\textwidth]{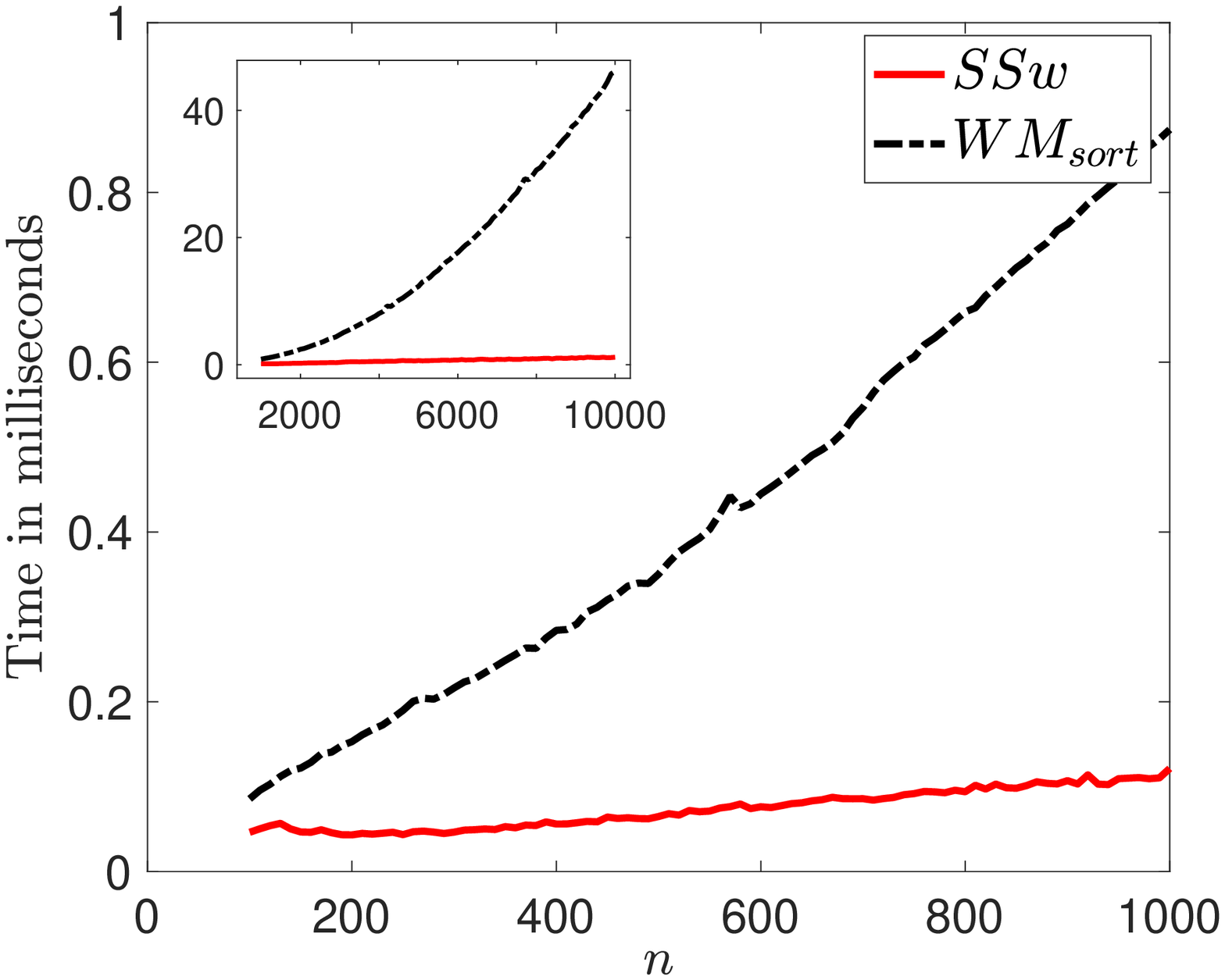}\hfill
\includegraphics[width=0.49\textwidth]{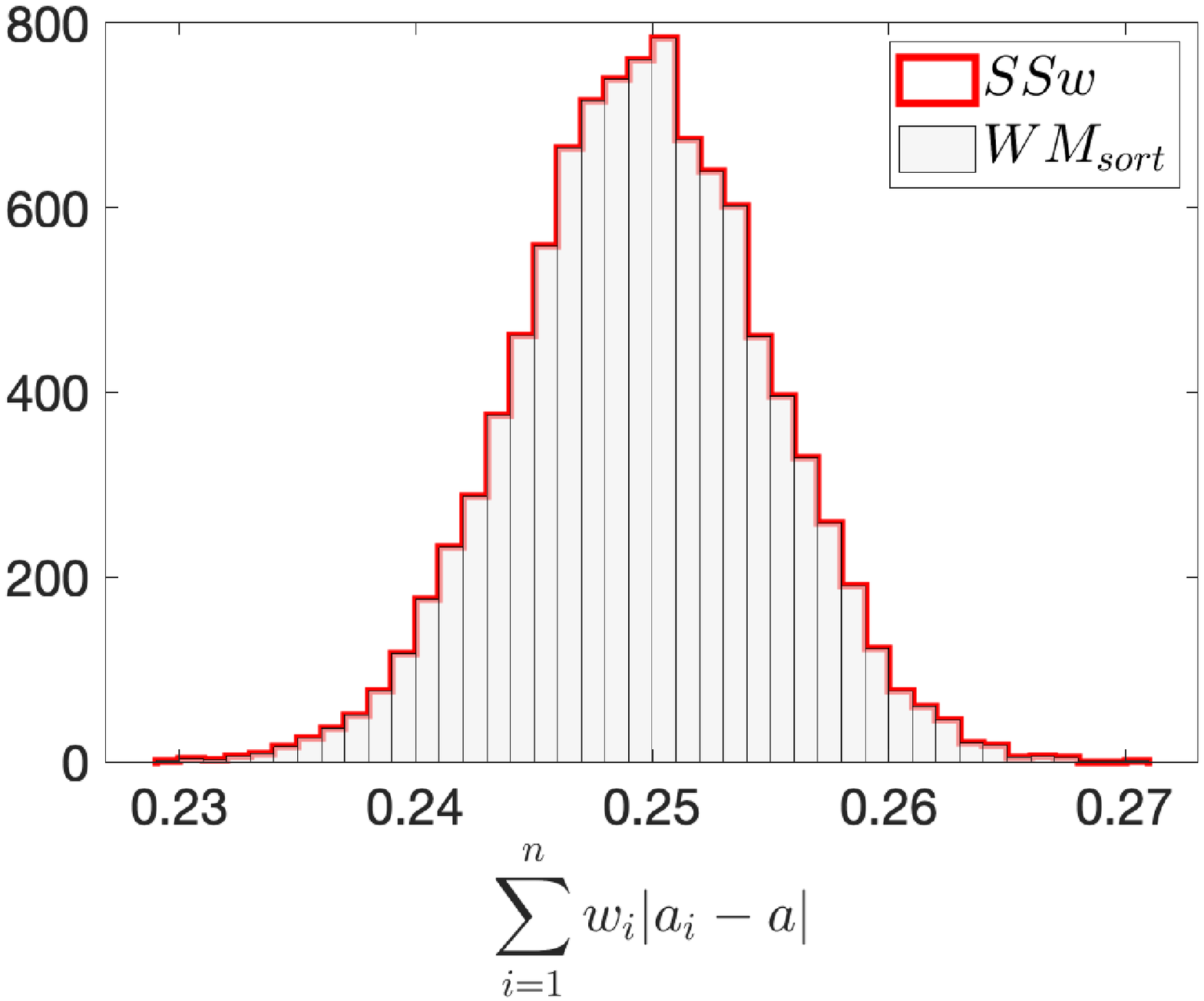} \\[3mm] 
\includegraphics[width=0.49\textwidth]{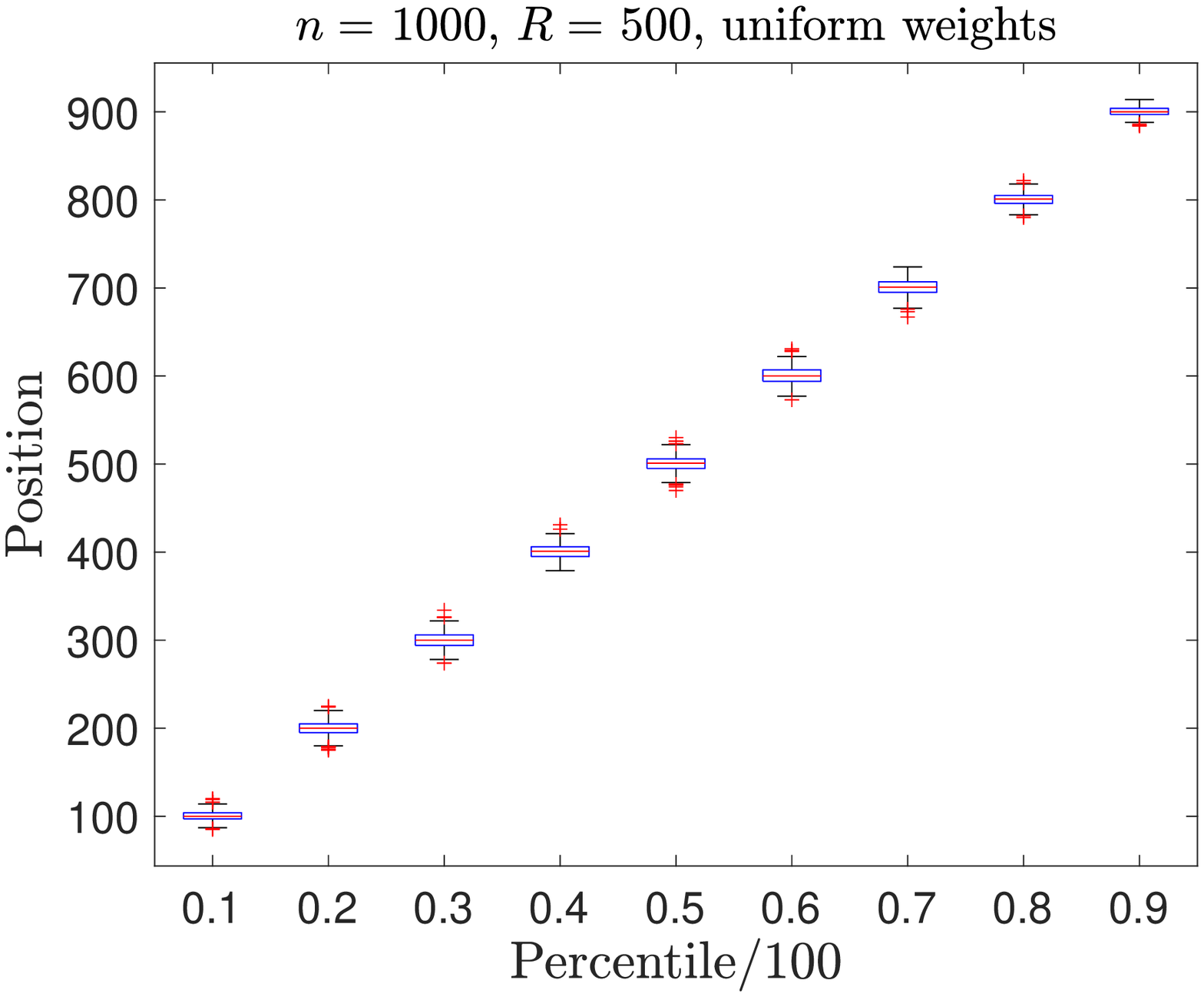}\hfill 
\begin{minipage}[b]{0.5\textwidth}
\caption{\label{WS:fig} 
Weighted \textsf{simpleselect} (\texttt{SSw}) and a $\mathcal{O}(n\log{}n)$ solution \citep{Haase:2022} (\texttt{WMsort}). %
Top-left: elapsed time for a weighted median, for $n\in [100 , 10000]$. 
Top-right: objective function values for weights and data from $U(0,1)$ and $n=1000$.
Bottom-left: positions $k^{*}_{p}$ returned  in \texttt{kstar} by \texttt{quickselctFSw} for various percentiles $100p$ and $n=1000$.
}
\end{minipage}
\end{figure}

The value $k^{*}$ of equation \ref{kstar:eq} is returned by function \texttt{quickselctFSw}  (listing \ref{DIVA_weighted:algo}, output variable \texttt{kstar}) with the input parameter $p=0.5$, that is the percentile $100p = 50\%$. 
More in general, for a generic percentile $100p \in [0,100]$, the function returns a value $k^{*}_{p}$ and a permutation of the weights which forces the sum of weight partition around $k^{*}_{p}$ to be as equal as possible and therefore the following difference as small as possible:
\begin{equation} 
\sum_{i=1}^{k^{*}_{p} - 1} w_{i} - \sum_{i=k^{*}_{p}+1}^{n} w_{i}.
\label{sum_weight_partition:eq}
\end{equation}
%
%
The histogram in the top-right panel of Figure \ref{WS:fig} shows that the resulting objective function values (\ref{weighted_median:eq}) are identical in \texttt{quickselctFSw} and an optimal $\mathcal{O}(n\log{}n)$ solution \citep{Haase:2022}, which sorts the weights for finding the smallest ones summing to half the total weights (respectively \texttt{SSw} and \texttt{WMsort} in the legend). 
The top-left panel of the same figure shows that our $\mathcal{O}(n)$ solution is advantageous even for small $n$, which is remarkable considering that it is obtained also here without compiling the code. 
The plot also shows that the $\mathcal{O}(n\log n)$ solution when $n$ is in the order of some thousands becomes very impractical.
Finally, the bottom-left panel shows that our extension to a generic weighted percentile works as expected, that is,  for uniformly distributed weights the positions $k^{*}_{p}$ returned by \texttt{quickselctFSw} are very close to the desired input percentile $100p$.

The code listing \ref{DIVA_weighted:algo} for \texttt{quickselctFSw} is obtained as a natural extension of \texttt{quickselctFS}. It is sufficient to embed the core part of the listing \ref{DIVA:algo} in a loop (starting at line 9) that checks the status of the sum or weights partition (\ref{sum_weight_partition:eq}) by extending the weighted median approach by  \cite{BleichOverton:83}. 
The core part of \texttt{quickselctFS} (lines 11-33) is called on the data assuming that the weighted percentile is in position $k=\lceil np \rceil$ (line 6).
This first iteration permutes vector D so that the weights at positions $(1:k-1)$ are smaller than the $k$th weight. At this point we check whether the array fulfills the  weights balance  (lines 36, 37, 41). If so, we stop iterating  (line 39). If not, we  apply again the core part of \texttt{quickselctFS} on either $D(1:k,:)$ with $k+1$, or $D(k:n,:)$ with $k-1$. The iteration will remove or add weights in order   to approach the optimal condition. 

The code spends most of the time - between 60\% and 70\%   -  in swapping rows (lines 19-20-21) within the loop at lines 16-25.
As the cell-pairs to swaps are expected to follow  the column major order of the \proglang{MATLAB} matrices, we traverse first the data elements \verb!D(:,1)!, which internally will be contiguous in memory, and then the weights elements \verb!D(:,2)!.  
This cell-wise swap approach is indeed found much faster than the conventional full-row swap of instructions
\begin{verbatim}
buffer=D(i,:);     D(i,:)=D(position,:);     D(position,:)=buffer;
\end{verbatim}
Finally note that in general the point that sum exactly to $100p\%$ of the total weight is between two data values. Therefore, at line 37  two values of $k$ can make \verb!D(k,2)! to satisfy equation (\ref{sum_weight_partition:eq}). 
We decided to solve the tie by returning the minimum between the two. In the case of the 50\% percentile, this is the \textit{lower weighted median}. This has to be taken into account when comparing \texttt{quickselctFSw} with solutions opting for the upper weighted median or an interpolant between the two, such as the mean.

\begin{figure}[t!]
\begin{center}
\begin{minipage}{0.85\textwidth}
\begin{lstlisting}[caption={Code distilled from \texttt{quickselectFSw.m}. Also available as R, \proglang{C} and C-mex files.\label{DIVA_weighted:algo} } ]
function [kD , kW , kstar]  = quickselectFSw(D,W,p)
% Extends SimpleSelect to weighted order statistics

n=length(D); left = 1; right = n; position=-1;
D=[D(:),W(:)]; % Values & weights go in pairs
k=ceil(n*p);   % Pivot index set to work in D(1:k,:)

BleichOverton = true;  %%% The external loop checks %%%
while BleichOverton    %%% the condition on weights %%%
    
    while (position~=k) % Internal loop is quickselectFS
        pivot      = D(k,:); 				 %| 
        D(k,:)     = D(right,:); 			 %| row swap
        D(right,:) = pivot; 				 %| 
        position   = left;
        for i=left:right
            if (D(i,1)<pivot(1,1))
                for s=1:2					 %|
                    buffer = D(i,s);		 %| cell-wise
                    D(i,s) = D(position,s);	 %| row swap
                    D(position,s) = buffer;  %| 
                end							 %|
                position = position+1;
            end
        end
        D(right,:)=D(position,:);
        D(position,:)=pivot;
        if  (position < k)
            left = position + 1;
        else
            right = position - 1;
        end
    end
    
    %%% Checks on weights - extends Bleich-Overton %%%
    Le=sum(D(1:k-1,2));
    if Le-p<=0 && p-Le-D(k,2)<=0 %% OK: stop computation
        kD=D(k,1); kW=D(k,2); kstar=k;
        BleichOverton=false;
    else             %% NOT OK: go back to quickselectFS 
        if  D(k,2)<2*(p-Le)  % Need to add a weight
            k=k+1; left=k; right=n;
        else                 % Need to remove a weight
            k=k-1; left=1; right=k;
        end
    end
    position=-1;
end
\end{lstlisting}
\end{minipage}
\end{center}
\end{figure}

\subsection{Application to the fast medcouple}
\label{medcouple:sec}

The weighted median is heavily used by a fast algorithm for the medcouple \citep{BrisHubertStruyf:2004}, a robust measure of skewness used to adjust the whiskers of a boxplot \citep{HubertVandervieren:2008} and avoid wrong declaration of outliers in asymmetric univariate data.
The \proglang{R} package robustbase (\url{https://cran.r-project.org/web/packages/robustbase})  and the \proglang{MATLAB} toolbox LIBRA (\url{http://wis.kuleuven.be/stat/robust.html}) implement the fast medcouple with wrapper functions to the same compiled \proglang{C} source,  to maximize speed.  
We have replicated faithfully the \proglang{C} function  \texttt{mlmc.c} in a pure \proglang{MATLAB} function,  \texttt{medcouple.m}, but we have replaced the computations of the median and weighted median with calls to our functions \texttt{quickselectFS} and \texttt{quickselectFSw}  or, for comparison, the $\mathcal{O}(n\log n)$ solution by \cite{Haase:2022}. 
For completeness, \texttt{medcouple.m} has been enriched with options to compute a ``naive'' $\mathcal{O}(n^2)$ solution and the quantile and octile approximations  used by \cite{BrisHubertStruyf:2004} for comparison. 
%
%
The consistency between \texttt{mlmc.c} in \proglang{C} and  \texttt{medcouple.m} in \proglang{MATLAB} has been ensured with systematic checks, ensuring same random numbers generation  with the Mersenne Twister framework. In Annex \ref{mtR:sec} we show how this is done and introduce a function to replicate random numbers also in \proglang{R}, which is not obvious. 

\begin{table}[tb!]
\centering
\begin{tabular}{|l|l|r|r|r|}
\hline 
&\textbf{Medcouple function}  & $\mathbf{n=10000}$ & $\mathbf{n=1000}$ & $\mathbf{n=100}$ \\ 
\hline 
1 & \texttt{mlmc.c} (full C-compiled) & 0.5272 & 0.0384 & 0.0114 \\ 
\hline 
  &\multicolumn{4}{|l|}{\texttt{medcouple.m} (full matlab-interpreted) with calls to:} \\
\hline 
2 & - \texttt{quickselectFSw} (mex) & 1.4456 & 0.0657 & 0.0170 \\ 
\hline 
3 & - \texttt{quickselectFSw} (jit) & 3.5361 & 0.1100 & 0.0155 \\ 
\hline 
4 & - \texttt{weightedMedian} (jit/sort) & 14.5733 & 0.2823 & 0.0272 \\ 
\hline 
5 & - \texttt{naive} (jit/sort) & 35.1412 & 0.1633 & 0.0073 \\ 
\hline 
\end{tabular} 
\caption{\label{mctime:tab}The medcouple with different weighted median solutions. Total time in seconds to apply the medcouple to 100 random samples of different size drawn from a log-normal distribution with function \texttt{lognrnd(0,1,n,1).}}
\end{table}
Table \ref{mctime:tab} gives an idea of the relative performance of the various solutions.
%
Our baseline (row 1) is the run with the fully-compiled \texttt{mlmc.c}, which is obviously advantaged. However, given that the algorithm spends most of the time in computing weighted medians, we expect comparable performances when \texttt{medcouple.m} uses the C-compiled mex of \texttt{quickselectFSw} (row 2).
The reported result is in accordance with expectancy, but note that this option could run faster by refining for speed the \proglang{C}-code of   \texttt{quickselectFSw} and replacing the loops in \texttt{medcouple.m} with vectorized code. 
Note finally that using the standard matlab function \texttt{quickselectFSw.m} (row 3) doubles the overall execution time (for medium-large $n$), yet keeping far from the time required by naive solutions based on sorting (rows 4 and 5): the need of avoiding them is obvious. 

\subsection{Application in digital filtering}
\label{filtering:sec}

\begin{figure}
\includegraphics[width=0.45\textwidth]{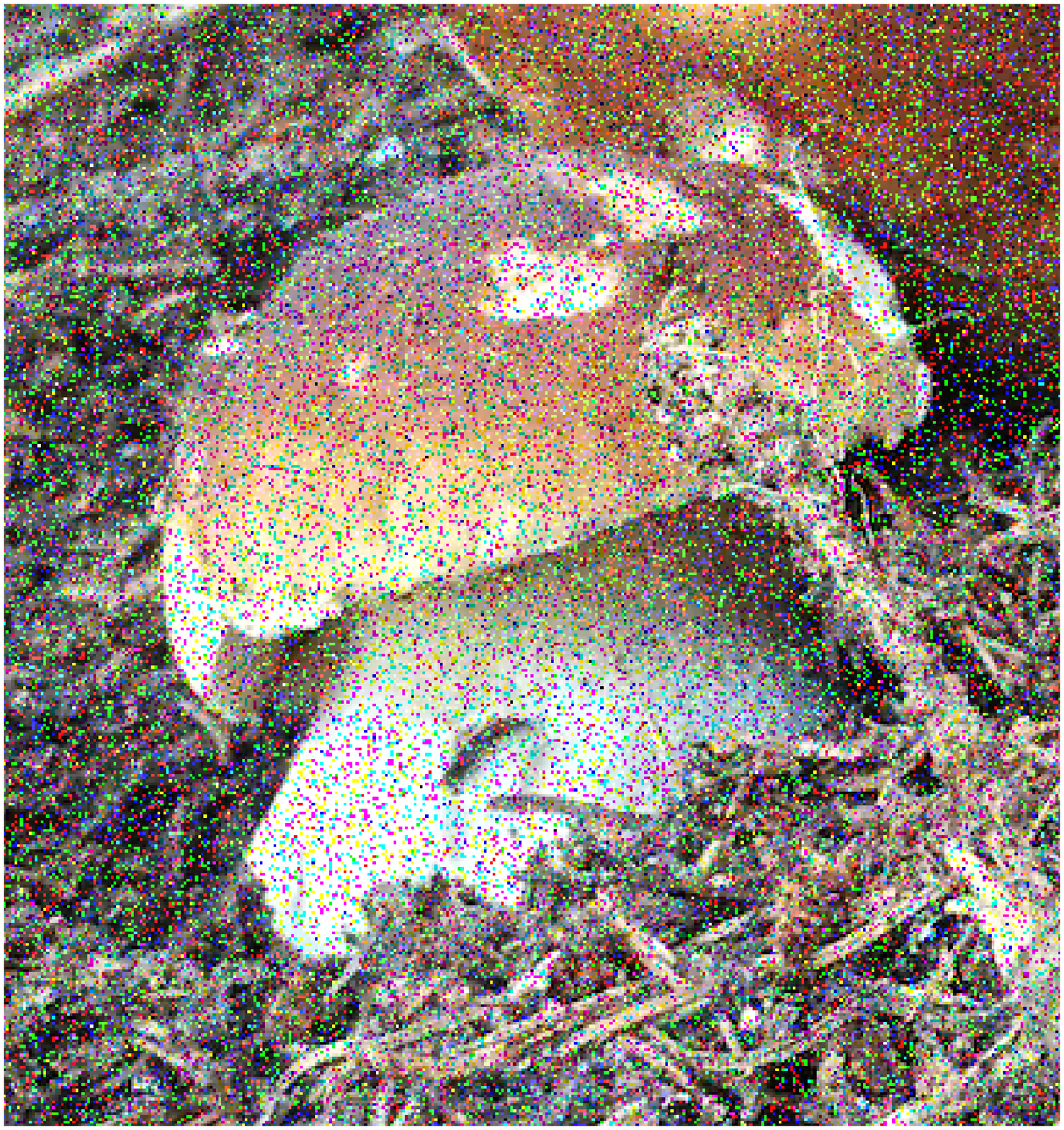}\hfill
\includegraphics[width=0.45\textwidth]{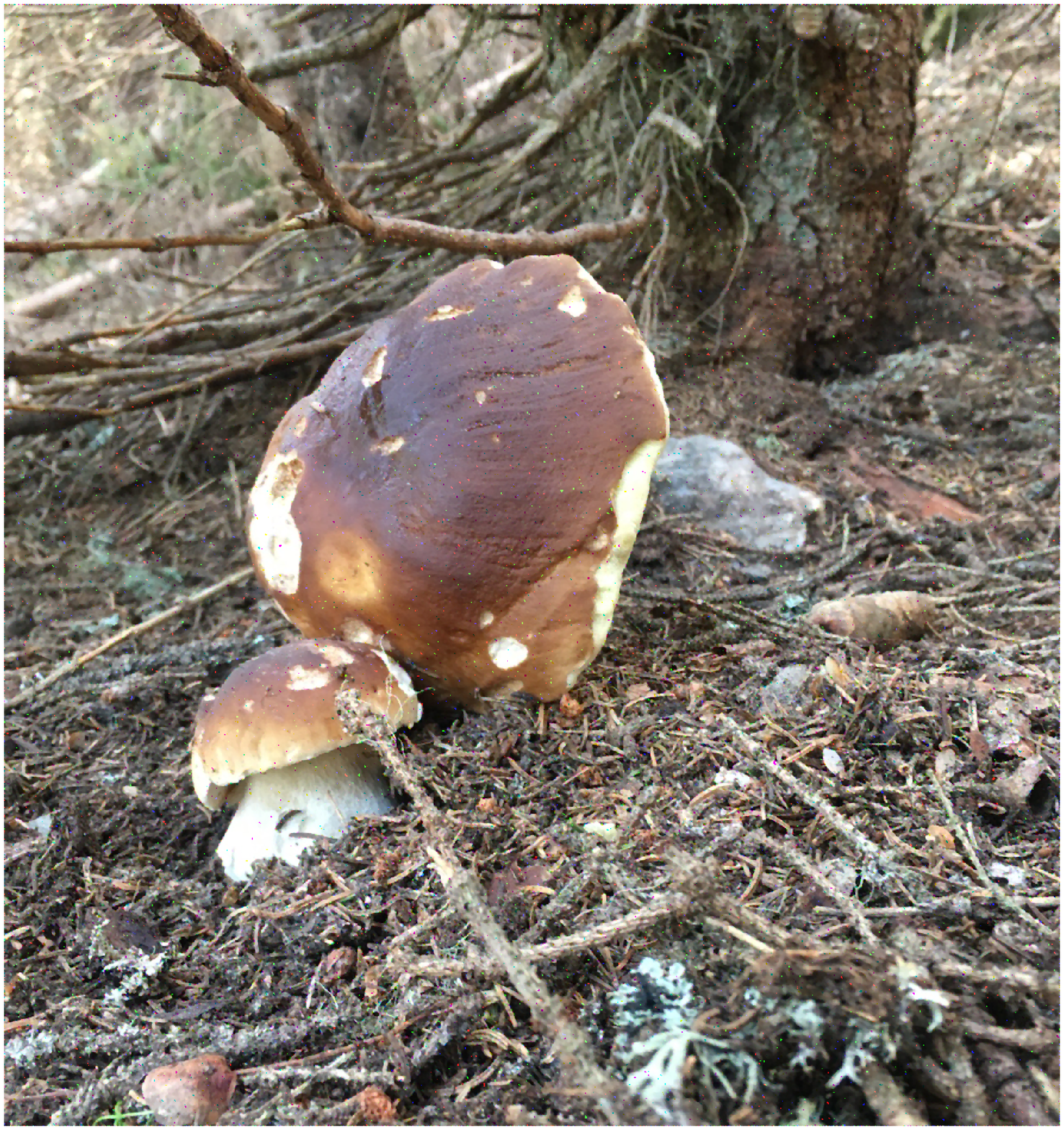}
\caption{\label{filtering:fig} 
A $1280 \times 1205 \times 3$ pixels photo denoised with the weighted median using a Wiener filter mask $W = [10,12,9 \,\,;\, 12,19,12 \,\,;\, 9,12,10]$. The detail on the left panel shows the added noise, which is removed by the filter (right panel).   
}
\end{figure}

Digital filtering covers many applications; here we take as an example the denoising of raster images, which was originally done by taking a number of values around a pixel and replacing it with the median of these values \citep{Pratt:1978}. The procedure is repeated within a window sliding throughout the image. \cite{Brownrigg:1984} has shown that the weighted median works better, especially to remove specific structural patterns. 
Obviously in this process the weighted median is used a very large number of times, depending on the image resolution. The listing \ref{denoising:code} shows that the calls to \texttt{quickselectFSw.m} would be $(row-3) \cdot (col-3) \cdot rgb = 4,604,862$ for a $3 \times 3$ weight mask and a $1280 \times 1205 \times 3$ photo like the one in Figure \ref{filtering:fig} 
where we added a certain percentage of gaussian noise. 
If we replace at line 36 the call to  \texttt{quickselectFSw.m} with the weighted median,
\vspace{-4mm}\begin{lstlisting}[frame=none , firstnumber=36] 
	[~ , Zclean(ind5)] = weightedMedian(A,W);
\end{lstlisting}
which relies on \texttt{sort}, the time execution raises more than 4 times (from 16 to 66 seconds in our test), which is a lot considering that in this case the weighted median is executed repeatedly on a small array of 9 values where the gain of replacing \texttt{sort} with \texttt{quickselectFSw} is in principle very limited. 

\begin{figure}[tb!]
\begin{center}
\begin{minipage}{0.85\textwidth}
\begin{lstlisting}[caption={Use of \texttt{quickselectFSw.m} to denoise a photo.\label{denoising:code} } ]
imfile = 'boletus1280.jpg';
pnoise = 0.2;   % percentage of noise to add

I = imread(imfile); % read image
%I = rgb2gray(I);   % uncomment for black and white
Z = double(I);
[row , col , rgb] = size(Z);

%% add noise to the image
x    = rand(size(Z));
d    = x < pnoise/2;
Z(d) = 0;    % Set to minimum value
d    = find(x >= pnoise/2 & x < pnoise);
Z(d) = 255;  % Set to saturated value

%% Weight-mask for the Wiener filter
W = [10,12,9; 12,19,12; 9,12,10];
W = W ./ sum(sum(W)); W = W(:);

%% denoise with weighted median
%  requires (col-3)*(row-3)*rgb calls to quickselectFSw
Zclean = Z(:); 
for qsw=0:1
    for y = 2:1:(col-1)*rgb
        for x = 2:1:row-1
            ind1 = x-1 + (y-1 - 1).*row;
            % = sub2ind([row,col],x-1,y-1);
            ind2 = x-1 + (y   - 1).*row;
            % = sub2ind([row,col],x-1,y);
            ind3 = x-1 + (y+1 - 1).*row;
            % = sub2ind([row,col],x-1,y+1);
            ind5 = x   + (y   - 1).*row;
            % = sub2ind([row,col],x,y);
            imask = [ind1; ind1+1; ind1+2; ind2; ind2+1; ind2+2; ind3; ind3+1; ind3+2];
            A = Z(imask);
            Zclean(ind5) = quickselectFSw(A,W,0.5);
        end
    end
end
Zclean = reshape(Zclean,size(Z));
figure; imshow(uint8(Zclean))
\end{lstlisting}
\end{minipage}
\end{center}
\end{figure}

\section{Conclusion}
\label{conclusion:sec}

Forty years ago \cite{BleichOverton:83} observed with a certain surprise that the best known linear algorithms for computing the median and weighted median  can take in practice more CPU time than the ``naive'' counterparts based on sorting, even for arrays of several thousands elements. 
This happens because of a constant but not negligible level of complexity linked to impractical data structures or problematic heap or stack memory management issues proper of intricate algorithms.  
It is therefore comprehensible that linear selection algorithms are hard to find in the main distribution of many statistical and programming environments.
%
As for weighted percentiles, to our knowledge the offer covers generally the weighted median and rarely with linear complexity solutions. 
We have shown that in robust statistics still today it can be convenient to resort to (weighted) selection algorithms and that our simplifications and generalizations  neutralise the constant computational overhead.   

The software necessary to replicate the results in the paper is available in \url{https://github.com/UniprJRC/FSDApapers}, under the folder \textsf{simpleselect} \texttt{/ArticleReplicabilityCodes}. 
We provide the key functions \texttt{quickselectFS}  and \texttt{quickselectFSw} as \proglang{MATLAB}, \proglang{R}, \proglang{C} and \proglang{C-mex} sources, the latter also compiled as binary \texttt{mex} files for the Linux, MacOsX and MS-Windows  platforms. 
These functions are also hosted by the standard distribution of FSDA, which can be downloaded from the GitHub space \url{UniprJRC} as well. 
The \proglang{R} community and users of other open environments like Python may run FSDA tools through the (automatically generated)  \proglang{C}-codes and the corresponding \proglang{R}-wrappers available in the  GitHub projects \texttt{FSDA-MATLAB\_Coder} and \texttt{fsdaR}  \citep{FSDA:toolbox}.


\appendix
\newpage
\section{Appendices}

\subsection{Expected increments of \texttt{position} in simpleselect}
\label{result1:sec}

In Listing \ref{DIVA:algo}, the `\textcolor{blue}{\texttt{for}} loop' is initialised with  \texttt{position=left=1} and \texttt{right=n} and receives a certain pivot value $a$ from the outer `\textcolor{blue}{\texttt{while}} loop'. Then,  $a$ is compared with all elements of $A$, assumed to originate from its same (unknown) parent distribution. 
Given that $a$ is taken from the element $a_k\in A$  (line 8), the pivot is actually compared with the $N=n-1$ elements of $A \setminus \{ a_k \}$  with the following $N+1$ possible outcomes:
\begin{equation}
 a \le a_{(1)} \quad ; \quad a_{(1)} < a \le a_{(2)} \quad ; \quad  \ldots  \quad ; \quad a_{(N-1)} < a \le a_{(N)} \quad ; \quad a > a_{(N)}
 \label{outcomes:eq}
\end{equation}
The last outcome obviously occurs when $a_{k} = a_{(n)}$. 
If we think about $A$ as one of $L$ independent samples analysed each by \textsf{simpleselect}, we can use a statistical extreme value problem formulated by \cite[Lecture 2, Plotting Positions]{gumbel:1954} on the cumulative distribution $F(a_{(m)})$ of the $m$-th smallest value among $N$ observations,  stating that:
\begin{equation}
 P(a \le a_{(m)}) = \lim_{L\rightarrow \infty} r_L(a \le a_{(m)}) = \lim_{L\rightarrow \infty}F(a_{m:L}) = E(F(a_{(m)})) = \frac{m}{N+1} 
  \label{gumbel:eq}
\end{equation}
where $r_L(\cdot)$ denotes the relative cumulative frequency computed on the $L$ samples and $a_{m:L}$ the set of  $L$ individual $m$th ranked values $a_{(m)}$.
Note that the cdf (\ref{gumbel:eq}) is a step function increasing by $\frac{1}{N+1}$ in each of the $N+1$ intervals representing the outcomes of (\ref{outcomes:eq}). This also means that $P(a=a_{(m)}) = \frac{1}{N+1} $ independently from $m$.
A compact demonstration of (\ref{gumbel:eq}) with discussion on related results, can be found in \cite{Makkonen:2008}.

Now, consider the subset of $A$ formed by the elements that satisfy the `\textcolor{blue}{\texttt{if}} statement' at line 18 and make \texttt{position} to increment by 1. We can denote the subset with $A_{lr} = \{a_i < a \; ; \; i=l,\ldots ,r \}$, $l$ and $r$ being respectively the indexes of the \texttt{left} and \texttt{right} pointers. 
The expected number of increments of \texttt{position} is the mean of the random variable  $X = |A_{l,r}|$, which we derive here for the first iteration of \textsf{simpleselect}, where:
\[X = |A_{1,n}| = t \quad \mbox{if} \; a=a_{(t+1)} \quad \mbox{for} \;  t=0,1,\ldots , N \]
%
%
%
Given that the cdf (\ref{gumbel:eq}) implies $\; P(X=t) = \frac{1}{N+1}$ independently from $t$, we have:
 \begin{equation}
E[X]  = \sum_{t=0}^{N} t \cdot P(X=t) = \frac{1}{N+1} \sum_{t=1}^{N} t = \frac{N (N+1)}{2 (N+1)} = \frac{N}{2}= \frac{n-1}{2}
 \label{expected_increments:eq}
 \end{equation}

\subsection{Implementation of the Vervaat perpetuities}
\label{vervaat:sec}

A perpetuity is a random variable of the form:
 \[ Y = W_1 + W_1 \cdot W_2 + W_1 \cdot W_2 \cdot W_3 + \ldots \]
 where the $W_i$ are an independent, identically distributed sequence of random variables. If each $W_i$ has the same distribution,  say $W_i \sim W$, then $Y \sim W(1 + Y)$ for $Y$ and $W$ independent.
The comparisons of Hoare's \texttt{Find} are distributed asymptotically as a particular perpetuity called Dickman, with $W \sim Unif([0,1])$. Unfortunately such distribution has no closed form.

The Dickman can be also seen as a special case of Vervaat perpetuity, which is such that $W_i \sim U^{1/\beta}$ for some $\beta \in (0,\infty)$ for $U \sim Unif([0,1])$. In other words, the Dickman distributon is a Vervaat perpetutiy with $\beta = 1$.
 A generalization of the perpetuity takes the form \[ Y =
 \sum_{n=0}^{\infty}A_n \prod_{i}^{n} W_i\] with $A_n$ not necessarily
 equal to $1$, which is known as Takacs distribution. 
Our implementation of the Vervaat family (\texttt{vervaatxdf.m}) (listing \ref{vervaat:algo}) follows \cite{bar+pra:19}, who introduced a feasible and elegant method for computing the probability density and distribution functions that avoids brute force simulation (code by the authors exists in Wolfram's Mathematica).  
For comparison we also ported from \proglang{R} to \proglang{MATLAB} the recursive simulation approach of \cite{cloud+huber:2017}, which is accurate but much slower (\texttt{vervaatsim.m}). 
For completeness, we implemented a function to simulate random variates from the Vervaat (\texttt{vervaatrnd.m}), using one of the two methods above.

\begin{center}
\begin{minipage}{0.85\textwidth}
\begin{lstlisting}[caption={Code distilled from the FSDA toolbox. Follows Wolfram's Mathematica code kindly shared with us by \cite{bar+pra:19}. User documentation: \protect\url{http://rosa.unipr.it/FSDA/vervaatsim.html}.  \label{vervaat:algo} } ]
function [f , F , x] = vervaatxdf(betav,nx,pascalM)
%pdf and cdf of a Vervaat perpetuity.
% betav  : Distribution parameter value. 
% nx     : Number of evaluation points. 
% pascalM: A precomputed Pascal matrix, used to speed up things in simulations. Remark: bc(a,b) = pascalM(a-b+1,b+1).

eulerGammaval = 0.577215664901532860606512090082402431;
n  = 100;
if nargin==0, betav = 1; end; % default is a Dickman
if nargin<2, nx = 1; end; 
if nargin<3, pascalM = pascal(n+1); end; 

h  = zeros(1,n);
L  = zeros(1,n+1);
ex = 5*betav;
x  = ex.*rand(1,nx); %nx random numbers from U in [0,ex]
f  = zeros(1,nx);
F  = zeros(1,nx);

for k=1:nx
    s=n/x(k);
    h(1) = betav * (exp(-s) - 1)/s;
    for j=2:n
        h(j) = (betav / s) * (-1)^(j-1) * exp(-s) - ((j-1)/s * h(j-1)) ;
    end
    L(1) = exp(betav * (- eulerGammaval - gammainc(0,s) - log(s) ));
    for i=2:n+1
        Li = 0;
        for j=0:i-2
            Li = Li + pascalM(i-2-j+1,j+1) * h(i-1-j) * L(j+1);
           %Li = Li + bc(i-2,j) * h(i-1-j) * L(j+1);
        end
        L(i) = Li;
    end
    % pdf
    f(k)  = ( (-1)^n / factorial(n) ) * s^(n+1) * L(n+1);
    % cdf
    F(k) = sum((-s).^(0:n) ./ factorial(0:n) .* L((0:n)+1));
end
\end{lstlisting}
\end{minipage}
\end{center}

\subsection{Call \texttt{quickselectFS.c} and \texttt{quickselectFSw.c} from Python}
\label{python:sec}

All statistical programming environments can integrate \proglang{C} functions to eliminate performance bottlenecks or compute specific algorithms. The typical approach is to write a code in the environment in use (the wrapper) that maps the original data types to those in the \proglang{C} code, initializes the function parameters and captures the result from the call to the compiled \proglang{C} function. 
The code listing \ref{python:diva} illustrates how \texttt{quickselectFS.c} can be called from Python. The call to \texttt{quickselectFSw.c} is similar.

\begin{center}
\begin{minipage}{0.85\textwidth}
\begin{lstlisting}[language=Python , caption={A wrapper calling \texttt{quickselectFS.c} from Python. \label{python:diva} } ]
# This Python script illustrates an easy way to call
# quickselectFS.c from Python, passing a Numpy array
# as the first argument.
#
# The .c source file should be compiled as follows:
#
# gcc -fPIC -shared -o simpleselect.so simpleselect.c 
#
# The C-compatible data types are built by the
# foreign Python function library ctypes. 

import ctypes
from numpy.ctypeslib import ndpointer
import numpy as np

lib = ctypes.cdll.LoadLibrary("simpleselect.so")
quickselectFS = lib.quickselectFS

quickselectFS.restype = ctypes.c_double
quickselectFS.argtypes = [ndpointer(ctypes.c_double, flags="C_CONTIGUOUS"),
                ctypes.c_int,
                ctypes.c_int]

A = np.arange(-5.0, 6.0, 1, dtype = float)
print(A)
np.random.shuffle(A)

# quickselectFS works in-place, so we have to create a copy of vector A if we want to retain the original A.

aux = A.copy()

k = 7
k_th_ord_stat = quickselectFS(aux, len(aux), k)

print(k_th_ord_stat)

del aux

\end{lstlisting}
\end{minipage}
\end{center}

\subsection{Generating same random numbers in MATLAB, C and R}
\label{mtR:sec}

To generate same random numbers in different languages just requires, in principle, the adoption of the same generation method. A famous one is  the Mersenne Twister algorithm by \cite{Mersenne-Twister:1998}, which is available for various languages at \url{http://www.math.sci.hiroshima-u.ac.jp/m-mat/MT/emt.html}. 
The code listing \ref{mtC:algo} illustrates how to use it in \proglang{C} to create from a given seed an array $D$ of uniform random numbers in $[0,1)$,  and apply on it \texttt{quickselectFS}. 
To replicate these numbers in \proglang{MATLAB} from the same seed is easy, as it includes since 2005 (R14SP3) built-in support for the Mersenne Twister \texttt{mt19937ar}. It is sufficient to run:
\vspace{-4mm}\begin{lstlisting}[frame=none , numbers=none, language=matlab] 	
RandStream.setGlobalStream(RandStream.create('mt19937ar','seed',896));
D = rand(n,1);
\end{lstlisting}
or equivalently
\vspace{-4mm}\begin{lstlisting}[frame=none , numbers=none, language=matlab] 	
myseed = 896;  rng(myseed , 'twister');
D = rand(n,1);
\end{lstlisting}
Note that instructions below produce the same $n$ integers between 1 and $N$
\vspace{-4mm}\begin{lstlisting}[frame=none , numbers=none , language=matlab] 	
D = ceil(rand(n,1)*N);  				D = randi(N,n,1);
\end{lstlisting}
that is, we can easily obtain uniform integers from uniform floats.
With \proglang{Python} the approach is similar, as it has the Mersenne Twister as core generator with the same underlying \proglang{C} library.

\begin{figure}[b!]
\begin{center}
\begin{minipage}{0.99\textwidth}
\begin{lstlisting}[caption={Generate same random numbers with Mersenne Twister: C-side. }  \label{mtC:algo} ]
#include <stdio.h>
#include <stdlib.h>
#include <stdint.h>
#include <math.h>
#include "simpleselect.h"	// quickselectFS/quickselectFSw
#include "mersenne.h"       // mt19937ar, Mersenne Twister 
							// with improved initialization
int main() {
    int n, k;
    printf("\n Choose sample size n: "); 
    scanf("%d",&n);
    printf("\n Choose order statistic k (<=n): ");
    scanf("%d",&k); 
    double D [n];
    
    uint32_t seed = 896;        // set a seed
    init_genrand(seed);         // initialise mt[n] with seed 
    for (int i=0; i<n; ++i) {   // generates random number in [0,1)
      D[i] = genrand_res53();   // with 53-bit resolution
    }                               
       
    double wE =quickselectFS(D,n,k-1); // apply quickselectFS
    printf("\n result: %f\n", wE);
    return 0;
}
\end{lstlisting}
\end{minipage}
\end{center}
\end{figure}

Instead, to replicate the same numbers in \proglang{R} is much more difficult, because the seeding algorithm in \proglang{R} does not follow exactly \texttt{mt19937ar}, using a different initialization and output transformation. 
This makes impossible, to our knowledge, to map directly \proglang{R} seed values to \proglang{MATLAB/C}'s. 
The \proglang{R} package \texttt{randtoolbox} offers an option aimed to generate random numbers along \texttt{mt19937ar}:
\vspace{-4mm}\begin{lstlisting}[frame=none , numbers=none,  language=R , backgroundcolor=\color{light-gray}] 	
set.generator("MersenneTwister", initialization="init2002", resolution=53, seed=myseed)
\end{lstlisting}
If, following the manual, we run it in \proglang{R} (V 4.0.5) with \texttt{myseed=12345}, we get 
\vspace{-4mm}\begin{lstlisting}[frame=none , numbers=none,  language=R , backgroundcolor=\color{light-gray}] 	
> runif(5)
[1] 0.1839188 0.2045603 0.5955447 0.6531771 0.2987037
\end{lstlisting}
Unfortunately, the manual reports a different sequence, which indeed we get in MATLAB with:
\vspace{-4mm}\begin{lstlisting}[frame=none , numbers=none,  language=MATLAB] 	
>> rand('twister', myseed); rand(1, 5)
ans =
    0.9296    0.3164    0.1839    0.2046    0.5677
\end{lstlisting}
It seems therefore that the behavior of \texttt{randtoolbox} is not stable. 
%
%

For this reason, we have integrated in FSDA a new function \texttt{mtR.m} that generates the same uniformly or normally distributed random numbers produced by the base \proglang{R} with \texttt{mt19937ar}. 
As it is not possible to map \proglang{R} seeds into \proglang{MATLAB}'s ones, the starting point of \texttt{mtR.m}  is the 626-element int32 vector containing the random number generator state used by \proglang{R} to generate random numbers, which we can get by executing:
\vspace{-4mm}\begin{lstlisting}[frame=none , numbers=none,  language=R , backgroundcolor=\color{light-gray}] 	
RNGkind("Mersenne Twister") # set "Mersenne Twister" "Inversion"
set.seed(myseed)
Rstate = .Random.seed
\end{lstlisting}
Of course, rather than generating in \proglang{R} a 626-element vector and pass it to \texttt{mtR.m} in \proglang{MATLAB}, it would be more convenient to build directly inside \proglang{MATLAB} the \proglang{R} state vector corresponding to a valid \proglang{R} seed. Fortunately, this is possible following the Mersenne Twister's \proglang{C} code, which we have introduced in \texttt{mtR.m} in the form of Listing \ref{mtRtoMstate:algo}.
\begin{figure}[tb!]
\begin{center}
\begin{minipage}{0.99\textwidth}
\begin{lstlisting}[caption={Generate same random numbers with Mersenne Twister: MATLAB-side. Sub-functions used by \texttt{mtR.m} to initialize the R state from a given seed. They mimic the Mersenne Twister's C code.}  \label{mtRtoMstate:algo} ]
function state = initMT_R(seed)
% Create an initial R Mersenne Twister state from a seed.
% c.f. RNG_Init() in RNG.c.
n = double(seed); % in case it was given as an int32
state = zeros(626,1,'int32');
for i = 1:50, n = int32lcg(n); end
for i = 2:626
    n = int32lcg(n);
    state(i) = n;
end
state(1) = 403;
state(2) = 624;
end

function n = int32lcg(n)
% Mimic the old glibc LCG that R uses for MT initialization, 
% in int32 arithmetic (MATLAB's int32 saturates, C's wraps).  
% c.f. RNG_Init() in RNG.c.
n = 69069 * n + 1;
if n >= 2^31 || n < -2^31
    n = mod(n+2^31,2^32) - 2^31;
end
end
\end{lstlisting}
\end{minipage}
\end{center}
\end{figure}

At this point, \texttt{mtR.m} can digest the generated \proglang{R} state by eliminating the first code, which stands for the RNG algorithm, reshuffling the state in the standard \proglang{MATLAB} form and recasting the vector from the signed integers used by \proglang{R} to the unsigned counterparts used by \proglang{MATLAB}; in short: 
\vspace{-4mm}\begin{lstlisting}[frame=none , numbers=none,  language=MATLAB] 	
Rstate      = initMT_R(myseed);
MATLABstate = [Rstate(3:end); Rstate(2)];
MATLABstate = typecast(int32(MATLABstate),'uint32');
\end{lstlisting}
Now, we can get \proglang{MATLAB}'s current global random number stream and set the state to that converted for \proglang{R} (or received directly from \proglang{R} if convenient):
\vspace{-4mm}\begin{lstlisting}[frame=none , numbers=none,  language=MATLAB] 	
Mstream       = RandStream.getGlobalStream();
Mstream.State = MATLABstate;
\end{lstlisting}
The last trick to keep in mind is to use the inverse transformation to compute a normal random variate (i.e. the standard normal inverse cumulative distribution function is applied to a uniform random variate). This replaces the \proglang{MATLAB} default, which is the \texttt{ziggurat} algorithm \citep{Marsaglia-Tsang:2000}. 

Function \texttt{mtR.m} contains numerous \proglang{MATLAB/R} examples, which can be used in simulation exercises involving both languages.

\newpage

\bibliographystyle{plainnat}
\bibliography{DIVA}

\end{document}